\documentclass[preprint]{aastex}
\usepackage{natbib}
\usepackage{hyperref}
\usepackage{epsfig}
\usepackage{amssymb,amsmath,lineno}
\usepackage{graphicx}
\usepackage{float}
\def\gsim{\mathrel {\vcenter {\baselineskip 0pt \kern 0pt
    \hbox{$>$} \kern 0pt \hbox{$\sim$} }}}

\begin{document}
\slugcomment{Published in ApJ as doi: 	10.1088/0004-637X/804/1/15}

\title{Searches for Anisotropies in the Arrival Directions \\of the 
 Highest Energy Cosmic Rays\\ Detected by the Pierre Auger Observatory}

\author{
{\bf The Pierre Auger Collaboration} \\
\begin{small}
A.~Aab$^{42}$, 
P.~Abreu$^{64}$, 
M.~Aglietta$^{53}$, 
E.J.~Ahn$^{81}$, 
I.~Al Samarai$^{29}$, 
I.F.M.~Albuquerque$^{17}$, 
I.~Allekotte$^{1}$, 
J.~Allen$^{84}$, 
P.~Allison$^{86}$, 
A.~Almela$^{11,\: 8}$
% ...
\\ \vspace{0.5cm}
$^{1}$ Centro At\'{o}mico Bariloche and Instituto Balseiro (CNEA-UNCuyo-CONICET), San 
Carlos de Bariloche, 
Argentina \\
% ...
$^{8}$ Instituto de Tecnolog\'{\i}as en Detecci\'{o}n y Astropart\'{\i}culas (CNEA, CONICET, UNSAM), 
Buenos Aires, 
Argentina \\
% ...
$^{11}$ Universidad Tecnol\'{o}gica Nacional - Facultad Regional Buenos Aires, Buenos Aires,
Argentina \\
% ...
$^{17}$ Universidade de S\~{a}o Paulo, Instituto de F\'{\i}sica, S\~{a}o Paulo, SP, 
Brazil \\
% ...
$^{29}$ Institut de Physique Nucl\'{e}aire d'Orsay (IPNO), Universit\'{e} Paris 11, CNRS-IN2P3, 
France \\
% ...
$^{42}$ Universit\"{a}t Siegen, Siegen, 
Germany \\
% ...
$^{53}$ Osservatorio Astrofisico di Torino  (INAF), Universit\`{a} di Torino and Sezione INFN, 
Torino, 
Italy \\
% ...
$^{64}$ Laborat\'{o}rio de Instrumenta\c{c}\~{a}o e F\'{\i}sica Experimental de Part\'{\i}culas - LIP and  
Instituto Superior T\'{e}cnico - IST, Universidade de Lisboa - UL, 
Portugal \\
% ...
$^{81}$ Fermilab, Batavia, IL, 
USA \\
% ...
$^{84}$ New York University, New York, NY, 
USA \\
% ...
$^{86}$ Ohio State University, Columbus, OH, 
USA \\
% ...
(\ddag) Deceased \\
(a) Now at Konan University \\
(b) Also at the Universidad Autonoma de Chiapas on leave of absence from Cinvestav \\
(d) Now at Unidad Profesional Interdisciplinaria de Ingenier\'{\i}a y Tecnolog\'{\i}as
Avanzadas del IPN, M\'{e}xico, D.F., M\'{e}xico \\
(e) Now at Universidad Aut\'{o}noma de Chiapas, Tuxtla Guti\'{e}rrez, Chiapas, M\'{e}xico \\
(f) Also at Vrije Universiteit Brussels, Belgium \\
\end{small}
}

\begin{abstract}
We analyze the distribution of arrival directions of ultra-high energy cosmic rays recorded at the Pierre Auger Observatory in 10 years of operation. The data set, about three times larger than that used in earlier studies, includes  arrival directions with zenith angles up to $80^\circ$, thus covering from $-90^\circ$  to $+45^\circ$ in declination. 
After updating the fraction of events correlating with the active galactic nuclei (AGNs) in the V\'eron-Cetty and V\'eron catalog, we subject the arrival directions of the data with energies in excess of 40 EeV to different tests for anisotropy. We search for  localized excess fluxes and for self-clustering of event directions at angular scales up to $30^\circ$ and for different threshold energies between 40~EeV and 80~EeV. We then look for correlations of cosmic rays with celestial structures both in the Galaxy (the Galactic Center and Galactic Plane) and in the local Universe (the Super-Galactic Plane). We also examine  their correlation with different populations of nearby extragalactic objects: galaxies in the 2MRS catalog, AGNs detected by Swift-BAT, radio galaxies with jets and the Centaurus~A galaxy.
None of the tests shows a statistically significant evidence of anisotropy. The strongest departures from isotropy (post-trial probability ${\sim}1.4$\%) are obtained for cosmic rays with $E>58$~EeV in rather large windows around Swift AGNs closer than 130~Mpc and brighter than $10^{44}$~erg/s (18$^\circ$ radius) and 
around the direction of Centaurus~A (15$^\circ$ radius).
\end{abstract}

\section{Introduction}
The measurements of the energy spectrum of ultra-high energy cosmic rays (UHECRs), their mass composition and the celestial distribution of their arrival directions serve in a complementary way to understand their origin. 
The acceleration mechanism as well as the propagation in the Galactic and intergalactic  media can be constrained by detailed studies of spectral features and of the evolution of the mass composition as a function of energy. 
In turn, and despite the fact that UHECRs are mostly charged particles, 
information on the sources might be 
contained in the distribution of their arrival directions, 
especially 
above a few tens of EeV where the magnetic deflections (at least of those cosmic rays with a small charge) may be of only a few degrees. 
A number of facts contribute to this expectation. Stringent limits to the flux of primary photons at such energies \citep{Augerpho} strongly constrain top-down models for the origin of UHECRs and hence favor astrophysical objects as accelerators. Also, at such energies the flux of cosmic rays is expected to be suppressed due to energy losses in their interactions with photons of the Cosmic Microwave Background (CMB) by the so-called GZK (Greisen-Zatsepin-Kuz'min) effect \citep{GZKg,GZKzk}. 
These interactions limit the distance from which a source can contribute to the flux at Earth.
For instance, this distance has to be less than $\sim 200$~Mpc for protons or Fe nuclei with energies above 60~EeV, 
and even smaller for intermediate mass nuclei \citep{horizon}. 
Thus, the number of candidate sources which could contribute to the measured fluxes at the highest energies is significantly reduced. 
Finally, the arrival directions of UHECRs are not expected to be completely isotropized by magnetic fields due to their very high rigidity.

A suppression in the flux of UHECRs at energies above 40~EeV has been established experimentally beyond any doubt  \citep{Hiressp,Augersp,TAsp}. The energy at which the spectrum steepens is in accordance with that expected from the GZK effect. However, this alone does not allow one to conclude whether the observed feature is due to propagation effects or to source properties, i.e., the maximum energy achievable in the  acceleration process. Information on the nature of UHECRs is one of the keys in discriminating between the two scenarios. The measurement of the cosmic ray composition has been addressed through the measurement of the depth of shower maximum, $X_\text{max}$ \citep{Augermass1,Augermass2,Hiresmass,TAmass}. Interpretations of Auger data through the most updated models of hadronic interactions \citep{Augerinter1,Augerinter2} indicate that the fraction of heavy nuclei increases above the energy of the ankle (the spectral hardening taking place at $E\simeq 5$~EeV) and up to the highest energies.  
However,  the small number of events does not allow one to probe the primary mass evolution in detail at energies in excess of 40 EeV, where there have been only 18 events available for the composition analysis.

To complement the spectrum and mass measurements, several studies of the distribution of arrival directions have been made with UHECR data. Using an early data set the Pierre Auger Collaboration reported evidence of anisotropy with a confidence level of 99\% in the distribution of cosmic rays with energy above about 57~EeV \citep{sc07,app08}. That analysis was based on the finding, through an {\it a-priori} test, of a correlation  within a small angular separation ($3.1^\circ$) between the UHECR arrival directions and the locations of nearby active galaxies (within 75~Mpc) in the V\'eron-Cetty and V\'eron (VCV) catalog \citep{vcv}. 
With an enlarged  data set the correlating fraction was found in later analyses to be lower \citep{app10,icrc11}, 
although still ${\sim}3\sigma$ above expectations from an isotropic distribution. 
Other tests on the data, using a variety of astronomical catalogs, yielded some further hints but no significant evidence of anisotropy \citep{app10}. 
It is interesting to note that both the Pierre Auger and the Telescope Array Collaborations have reported, although with a limited significance, concentrations of very high energy events in  regions of the sky of ${\sim}20^\circ$ radius,  
namely for $18^\circ$ around the radio galaxy Centaurus~A (Cen~A) in the case of Auger  \citep{app10} and in a $20^\circ$ radius window at declination  $\delta=43^\circ$  in the case of the Telescope Array \citep{TAhotspot}. Note that the hot spot observed around Cen~A is outside the field of view of the Telescope Array, while the one observed by the Telescope Array is only partially inside the field of view of the Auger Observatory when highly-inclined events are considered.

In the present situation  where the origin of the suppression in the flux of  the UHECRs  has not yet been understood, their mass composition is not precisely  known and the predictions of their deflections in magnetic fields are uncertain (also due to uncertainties in models of magnetic fields, see for example \citet{FarrarCC2014} for a recent review), a large number of events is essential in looking for anisotropies in a sky map. Whatever the origin of the suppression in their flux and whatever their nature, UHECRs are still expected to come from sources relatively close to the Earth where the galaxies are distributed non uniformly. 
Even if low-charge particles were to contribute only a fraction of the
primary cosmic rays, anisotropic signals on small angular scales may show up as the number of events gathered increases.
 In turn, should the UHECRs be significantly deflected, either due to their large charge or due to the presence of strong intervening magnetic fields, directional excesses might still be found at larger angular scales. Searches for such anisotropies have been made so far with data sets including a few dozen cosmic rays (for instance in \citet{app10} we published the arrival directions and energies of 69 events above 55~EeV and zenith angle $\theta\leq 60^\circ$, corresponding to an exposure of 20,370 km$^2$~sr~yr). In this paper we present a study of the arrival directions of UHECRs detected by the Pierre Auger Observatory in more than 10 years of data taking, with an exposure of about 66,000~km$^2$~sr~yr. The data set, including more than 600 events above 40~EeV, is described in Section~2. 
By including for the first time cosmic rays with zenith angles up to $80^\circ$, 
the field of view of the Auger Observatory has been extended to cover from $-90^\circ$ to $+45^\circ$ in declination. 

In the later sections we analyze the distribution of the arrival directions. In Section~3 we update the fraction of events correlating with AGNs in the VCV catalog.
In spite of the large data set (three times larger than that used in \citet{app10}), this test does not substantiate the initial evidence of anisotropy at energies larger than 53~EeV\footnote{This threshold was 57~EeV in the original calibration used in \citet{sc07,app08}. It became 55~EeV with the updated reconstruction used in \citet{app10}, corresponding to approximately 53~EeV  in the new energy scale considered in the present work (see Section 2).}. We consequently explore in the later sections the set of arrival directions for cosmic rays observed with energies above 40~EeV.
Since this energy corresponds to the onset of the suppression in the observed flux, we expect a limited number of contributing sources above such a threshold. 
Also, above this energy the angular deflections caused by intervening magnetic fields are expected to be of the order of a few degrees for protons, and $Z$ times larger in the case of nuclei with atomic number $Z$. 
We perform various tests to search for anisotropies in the data set, exploring a wide range of angular windows between $1^\circ$ and $30^\circ$ and energy thresholds from 40~EeV up to 80~EeV. The angular range is motivated, at the lower end, by the angular resolution of the measurement of the arrival directions  and, at the higher end, by the large deflections expected if cosmic rays are high-$Z$ nuclei. 
Considering energy thresholds higher than 40~EeV may help because it may involve smaller deflections and smaller GZK horizons, with the upper value of 80~EeV  still allowing  to have a sizeable number of events (22) in the analysis.
In Section~4 we study ``intrinsic'' anisotropies as can be revealed by the search for localized excesses of events over the exposed sky and by the analysis of the autocorrelation of arrival directions. 
In Section~5 we search for correlations with known astrophysical structures, such as the Galactic and Super-Galactic Planes and the Galactic Center. We study the cross-correlation with astrophysical objects that could be considered as plausible candidates for UHECR sources in Section~6. Specifically, we exploit flux-limited catalogs of galaxies (2MRS), of AGNs observed in X-rays (Swift BAT-70) and of radio galaxies with jets. For the last two samples, we perform an additional study, considering different thresholds in the AGN intrinsic luminosity. 
Finally, in Section~7 we focus on the distribution of events around the direction of Centaurus~A. After summarizing the main results in Section~8 we report in the Appendix the list of arrival directions and energies of the 231 UHECRs with energies above 52~EeV detected by the Pierre Auger Observatory between 2004 January 1 and 2014 March 31\footnote{The list of the events is available also at \url{http://www.auger.org/data/AugerUHECR2014.txt}.}.

\section{The Data Set}

The Pierre Auger Observatory \citep{nim04} is located in Malarg\"ue, Argentina, at latitude 35.2$^\circ$ S, longitude 69.5$^\circ$ W and an average altitude of 1400 m a.s.l. It comprises a surface detector (SD) made up of an array of water-Cherenkov stations overlooked by an air-fluorescence detector (FD) comprising a total of 27 telescopes at four sites on the perimeter of the array. The array consists of 1660 water-Cherenkov stations covering an area of about 3000 km$^2$.  
The SD samples the particle components of extensive air showers (mainly muons, electrons, and photons) with a duty cycle of nearly 100\%.

The data set analyzed here includes cosmic rays with energy above 40 EeV recorded by the SD from 2004 January 1 up to 2014 March 31. In earlier analyses of the arrival directions we have used events with zenith angles less than 60$^\circ$ (referred to as \textit{vertical}). Here we include, for the first time, those with zenith angles from $60^\circ$ up to 80$^\circ$ (dubbed \textit{inclined}). Selection, reconstruction and energy determination are different for the two event sets. The main characteristics of the data sets, including energy and angular resolution, are outlined below and details can be found in \citet{ave2007,HAS2014}.

Vertical events are accepted if at least four of the closest stations to the one with the highest signal are operational at the time of the event. 
We also require that the reconstructed shower core be contained within a triangle of operational stations, either equilateral or isosceles, of contiguous stations. 
This event selection, a less stringent one than that used in earlier works (where five operational neighboring stations were required), has been carefully studied using data. 
It ensures an accurate event reconstruction given the large multiplicity of triggered detectors 
(on average more than 14 stations are triggered in events with energy above 40 EeV). 
It also allows us to increase the number of vertical events by about 14\% in the period considered,  a value which is consitent with the increase in aperture gained with the more relaxed trigger.
On the other hand,  for inclined events 
we require that at least five active stations surround the station closest to the core position.
Given the large footprint of inclined showers on the ground (the average station multiplicity is larger than 30), such a fiducial criterion guarantees adequate containment inside the array. The described selections lead to 454 vertical and 148 inclined events with $E\geq 40$~EeV.  

The trigger and selection efficiency is 100\% for energies above 3 (4) EeV for vertical (inclined) showers. The exposure is consequently determined by purely geometrical considerations \citep{nim10,sc13} in both cases and for the period considered here it amounts to 51,753~km$^2$~sr~yr and 14,699~km$^2$~sr~yr, for the vertical and inclined samples, respectively.

For both data sets the arrival directions of cosmic rays are determined from the relative arrival times of the shower front in the triggered stations. 
The angular resolution, defined as the radius
around the true cosmic ray direction that would contain 68\% of the reconstructed shower directions, is better than 0.9$^\circ$ for energies above 10~EeV \citep{ar09}.

The ground parameters used to estimate the primary energy are different for the two data sets. 
The estimator for the primary energy of vertical showers is the reconstructed signal at 1000~m from the shower axis, denoted $S(1000)$. 
The energy reconstruction of an inclined shower is based on the muon content, denoted $N_{19}$, relative to a simulated proton shower with energy $10^{19}$ eV. 
In both cases, the energy estimators are calibrated using hybrid events (detected simultaneously by SD and FD) and using the quasi-calorimetric energy determination obtained with the air fluorescence detector \citep{sc13,HAS2014}. 
The statistical uncertainty in the energy determination is smaller than 12\% for energies above 10~EeV \citep{HAS2014,pe11}. 
The systematic uncertainty in the absolute energy scale, common to the two data sets, is 14\%.
The Pierre Auger Collaboration has updated the energy scale in \citet{Verzi2013} accounting for recent measurements of the fluorescence yield \citep{airfly}, a better estimate of the invisible energy \citep{tu13}, a deeper understanding of the detector, and an improved event reconstruction. 
The energy threshold of 55~EeV used in our previous publication \citep{app10} corresponds now to approximately 53~EeV with the new energy scale.

We note that the relative number of vertical and inclined events above 40~EeV, $454/148~\simeq~3.07\pm 0.29$, is consistent in view of the Poissonian fluctuations with the corresponding ratio of exposures, $51,753/14,699 \simeq 3.52$. On the other hand, the 14\% difference between  these ratios could also result from a ${\sim}4$\% mismatch between the vertical and inclined energy calibrations, which is compatible with the uncorrelated systematic uncertainties on the energy scale.

\section{Note on the Anisotropy Test with the VCV Catalog}

One of the anisotropy tests performed in our previous works was based on the V\'eron-Cetty and V\'eron catalog of active galactic nuclei \citep{vcv}. In an initial study we considered vertical events with $E\geq 40$~EeV collected from 2004 January 1 to 2006 May 26 (Period I).  We performed an exploratory scan over the energy threshold of the events, their angular separation from AGNs and the maximum AGN redshift. 
We found that the most significant excess appeared in the correlation of events with energy above 57~EeV and lying within $3.1^\circ$ of those AGNs closer than 75~Mpc. 
These parameters were then used for a search on independent data where it was found that 8 out of 13 events correlated, while 2.7 events (i.e., 21\% of the total) were expected to correlate by chance for an isotropic distribution of arrival directions. This finding had a probability of  $1.7{\times} 10^{-3}$ of happening by chance \citep{sc07,app08}. Subsequent analyses with enlarged statistics yielded a correlation still above isotropic expectations but with a smaller strength and essentially dominated by the initial excess. The level of correlation was $(38^{+7}_{-6})$\% in \citet{app10} and $(33\pm 5)$\% in \citet{icrc11}. 

Here we update this analysis, for historical reasons, by using the vertical data set described in Section~2 and the VCV catalogue used in \citet{sc07}. Excluding Period I, there are 146 events above 53~EeV: 41 events correlate with VCV AGNs, with the angular and distance parameters fixed by the exploratory scan.  The updated fraction of correlations is then (28.1$^{{}+3.8}_{{}-3.6})$\%, which is 2~standard deviations above the isotropic expectation of 21\%. 
On the other hand, note that since the VCV correlations involve many different regions of the sky (besides the fact that CRs with different energies have significant time delays), an explanation of the reduced correlation found after 2007 in terms of a transient nature of the signal  would not be natural. Hence, the high level of correlation found initially was probably affected by a  statistical fluctuation. We conclude that this particular test does not yield a significant indication of anisotropy with the present data set.

\section{General Anisotropy Tests}

\subsection{Search for a Localized Excess Flux over the Exposed Sky}

A direct  analysis of cosmic ray arrival directions is the blind search for excesses of events over the visible sky. To this aim we sample the exposed sky using  circular windows with radii varying from 1$^\circ$ up to 30$^\circ$, in 1$^\circ$ steps. The centers of the windows are taken on a 1$^\circ{\times} 1^\circ$ grid. 
The energy threshold of the events used to build the maps is varied from 40~EeV up to 80~EeV in steps of 1~EeV. To detect an excess we compare, for every window and  energy threshold, the number of observed events, $n_\text{obs}$, with that expected from an isotropic flux of cosmic rays, $n_\text{exp}$. The expected number of events for an isotropic distribution is obtained,  for each sky direction,  by numerically integrating the geometric exposures in the corresponding windows. We use the total number of vertical and inclined events to normalize the relative exposures of the two samples. Note that since the triggering is different in the two cases, this fraction is non-trivial.

\begin{figure}[H]
\centerline{\includegraphics[height=1.5in,angle=-0]{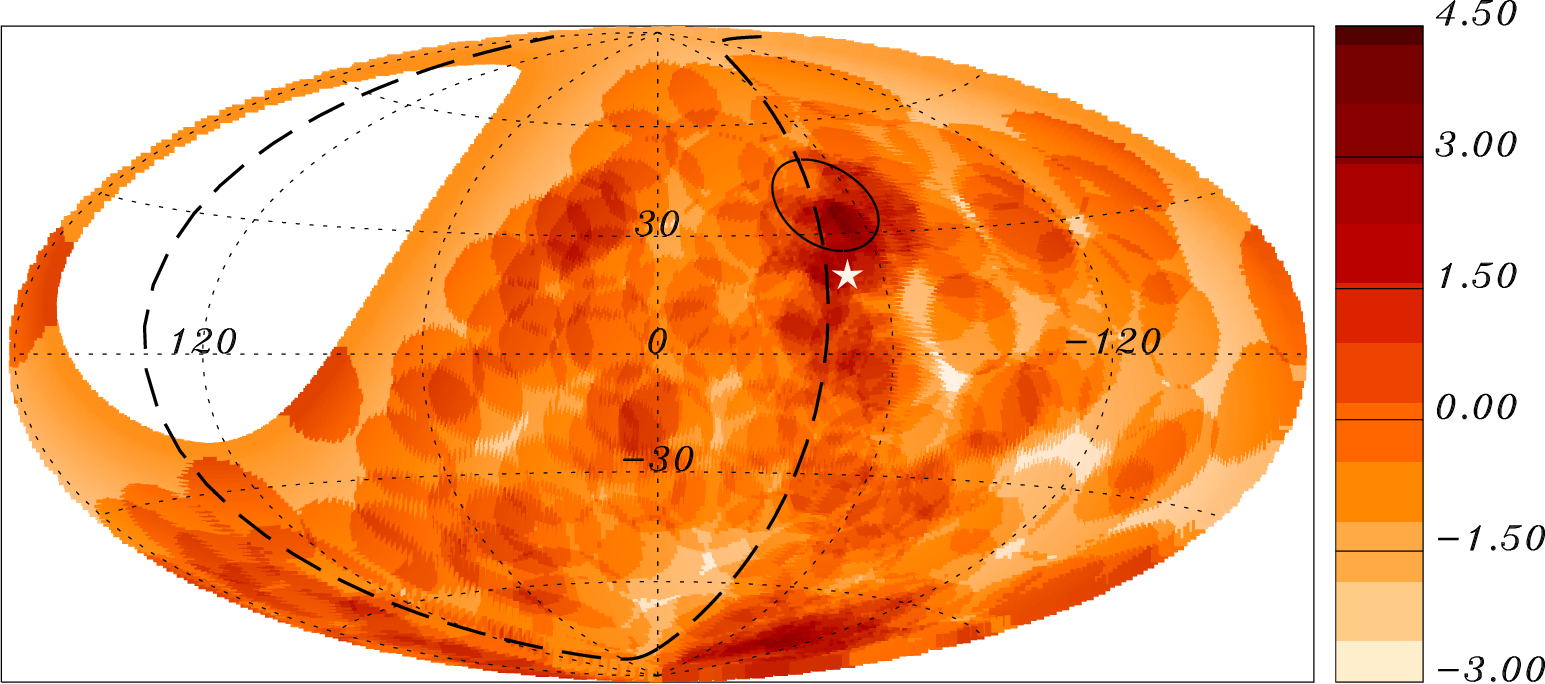}}
\vskip 1.0 truecm
\caption{Map in Galactic coordinates of the Li-Ma significances of overdensities in 12$^\circ$-radius windows for the events with $E\geq 54$~EeV. Also indicated are the Super-Galactic Plane (dashed line) and Centaurus~A (white star).} 
\label{f.wild}
\end{figure}

For each window we calculate the binomial probability $p$ of observing by chance in an isotropic flux an equal to, or larger number of events than that found in the data. We find the minimum probability, $p=5.9{\times} 10^{-6}$, at an energy threshold of 54~EeV and in a 12$^\circ$-radius window centered at right ascension and declination $(\alpha,\delta)=(198^\circ, -25^\circ)$, i.e., for Galactic longitude and latitude $(\ell,b)=(-51.1^\circ, 37.6^\circ)$, for which $n_\text{obs}/n_\text{exp}=14/3.23$. 
The map  of the Li-Ma \citep{lima} significances of the excesses of events with $E\geq 54$~EeV in windows of 12$^\circ$ radius is shown in Figure~\ref{f.wild}. The highest significance region just discussed, having a Li-Ma significance of 4.3$\sigma$, is indicated with a black circle. It is close to the  Super-Galactic Plane, indicated with a dashed line, and  centered at about $18^\circ$ from the direction of Centaurus~A, indicated with a white star. One should keep in mind that although the effects of a turbulent magnetic field would be just to spread a signal around the direction towards the source, a regular field coherent over large scales would give rise to a shift of the excess in a direction orthogonal to that of the magnetic field,  the size of both effects being energy dependent.

To assess the significance of this excess we simulated 10,000 sets of isotropic arrival directions containing the same number of events as the data set. In doing so, we keep the original energies of the events and assign to them random arrival directions according to the geometric exposure,  choosing randomly between vertical and inclined events according to their relative exposures. We apply to the simulated sets the same scans in angle and energy as those applied to the data. 
We find that values smaller than $p=5.9{\times} 10^{-6}$ 
 are obtained in 69\% of isotropic simulations and hence the excess found in the data turns out to be compatible with the maximum excesses expected in isotropic simulations. We note that in the region of the hot-spot reported by the Telescope Array Collaboration \citep{TAhotspot}, a 20$^\circ$ radius circular window centered at $(\alpha,\delta)=(146.7^\circ, 43.2^\circ$), which is partially outside our field of view, we expect to see 0.97 events with $E>53$~EeV if the distribution were isotropic and 1 event is observed.

\subsection{The Autocorrelation of Events}

Another simple way to test the clustering of arrival directions is through an autocorrelation analysis, which is particularly useful when several sources lead to excesses around them on a similar angular scale.
With this method one looks for excesses in the number of pairs of events, i.e., excesses of ``self-clustering,'' namely, we count the number of pairs of events, $N_\text{p}(\psi,E_\text{th})$, above a given energy threshold $E_\text{th}$ that are within a certain angular distance $\psi$. We do this at different energy thresholds, from 40~EeV up to 80~EeV (in steps of 1~EeV) and we look at angular scales from $1^\circ$ up to $30^\circ$ (in steps of 0.25$^\circ$ up to 5$^\circ$, and of 1$^\circ$ for larger angles). To identify an excess we compare the observed number of pairs with that expected from an isotropic distribution having the same number of arrival directions above the corresponding energy threshold.  For each energy threshold and angle we then calculate the fraction of isotropic simulations having an equal number to, or more pairs than the data, $f(\psi,E_\text{th})$. 
\begin{figure}[H]
\centerline{\includegraphics[height=2.6in,angle=-0]{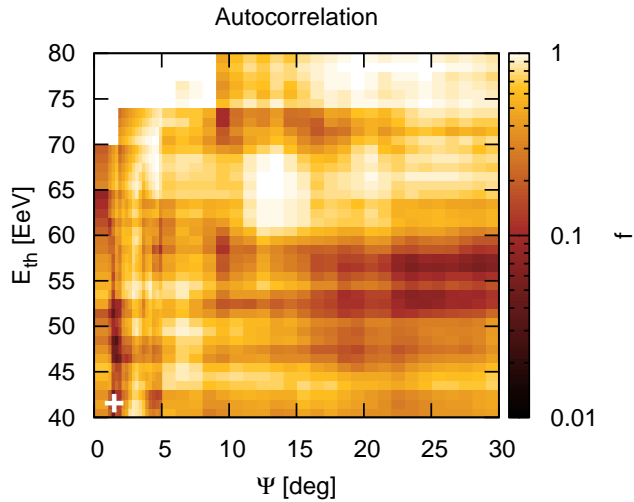}}
\vskip 1.0 truecm
\caption{Fraction $f$ obtained in the autocorrelation of events versus  $\psi$ and $E_\text{th}$.} 
\label{f.autocor}
\end{figure}
The result  is shown in Figure~\ref{f.autocor}  as a function of the angular distance and the energy threshold. The color code indicates the values obtained for $f$. The white cross corresponds to the parameter values leading to the minimum value of this fraction, $f_\text{min}=0.027$, which happens for $\psi=1.5^\circ$ and $E_\text{th}=42$~EeV. For these parameters, 30 pairs are expected on average for isotropic simulations while 41 are observed in the data. We calculate the post-trial probability for this excess, $P$, as the fraction of isotropic simulations which under a similar scan over $E_\text{th}$ and $\psi$ lead to a value of $f_\text{min}$ smaller than the one obtained with the data. The resulting value, $P\simeq 70$\%, indicates that  the autocorrelation is compatible with the  expectations from an isotropic distribution of arrival directions. 

\section{Search for Correlations with the Galaxy  and with the Super-Galactic Plane}

In the previous section we tested the intrinsic distribution of arrival directions of UHECRs, i.e., without formulating any hypothesis on the distribution of their sources. 
In the following
we consider specific astrophysical structures and objects as candidate sources. 
In this section we search for correlations with the Galactic and the Super-Galactic Planes as well as with the Galactic Center. On the one hand, a Galactic origin of UHECRs might give rise to an excess of arrival directions near the plane of the Galaxy, especially if a low-$Z$ primary component contributes to the CR flux. On the other hand, nearby galaxies (within 100~Mpc) show a clustering along the so-called Super-Galactic Plane, which contains several prominent (super) clusters such as Virgo, Centaurus, Norma, Pavo-Indus, Perseus-Pisces, Coma, etc., and hence extragalactic cosmic rays could be clustered near the Super-Galactic Plane.

\begin{figure}[t]
\centerline{
\includegraphics[height=2.2in,angle=0]{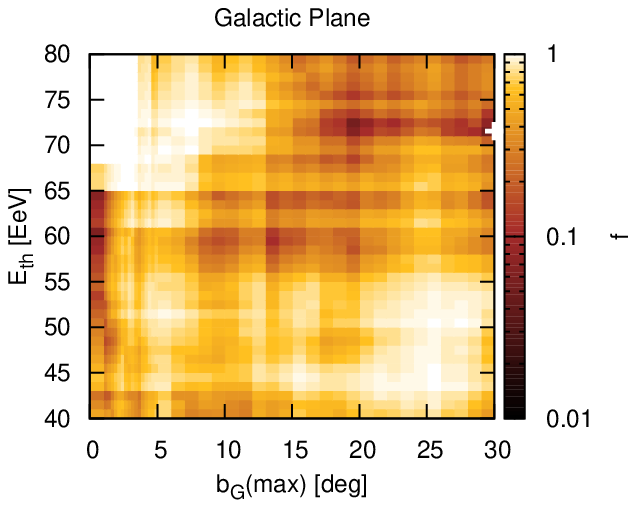}\ \ \ 
\includegraphics[height=2.2in,angle=0]{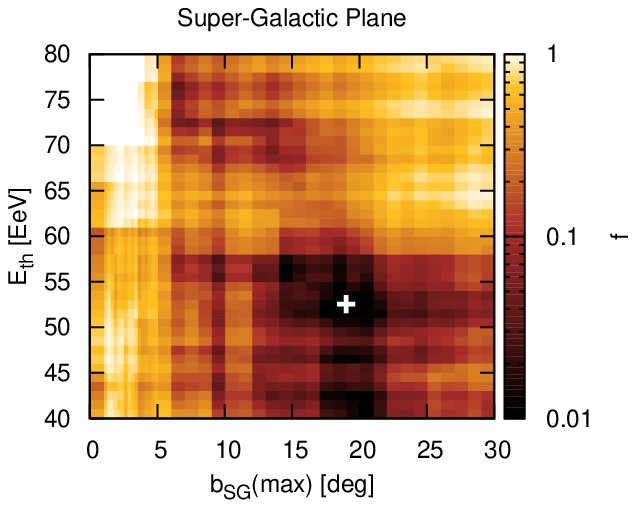}}
\caption{Fraction $f$  as a function of Galactic (left) or Super-Galactic (right) latitude band half width considered, for events with energies above $E_\text{th}$.} 
\label{f.gsg}
\end{figure}

\begin{figure}[t]
\centerline{
\includegraphics[height=2.3in,angle=0]{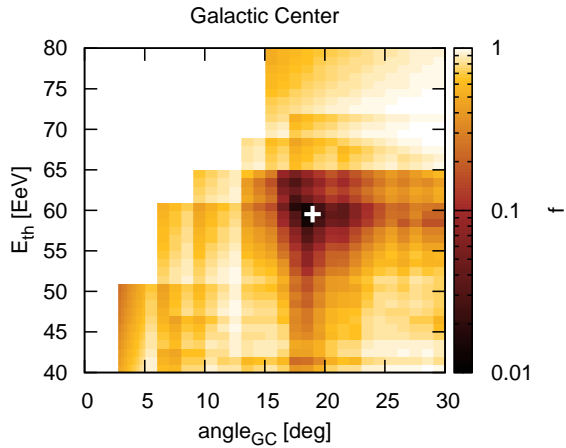}}
\caption{Fraction $f$  in circular windows around the Galactic Center as a function of the angular radius of the  window and $E_\text{th}$.} 
\label{f.gc}
\end{figure}

 We search for excesses of events as a function of Galactic (Super-Galactic) latitude, $b_\text{G}$ ($b_\text{SG}$), considering different latitude bands, $|b|<b(\text{max})$, with $b(\text{max})$ indicating the half width of the band.
To identify an excess we compare the number of events observed  within the latitude band considered with those obtained in  isotropic simulations for the distribution of arrival directions.  The plots in Figure~\ref{f.gsg}  display the fraction $f$ of isotropic simulations leading a larger number of events than the data   for the different energy thresholds and angular scales considered. The left figure is for the excesses in different latitude bands around the Galactic Plane, leading to a minimum value $f_\text{min}\simeq 0.05$ for $E\geq 72$~EeV and $b_\text{G}(\text{max})=30^\circ$, indicated with a white cross in the figure, in which case 29 events are observed while 22.8 would be expected on average in isotropic simulations. The penalised probability for obtaining a smaller value of $f_{min}$ in isotropic simulations after a similar scan is $P=70\%$. The right plot  is similar but for the excesses in different super-galactic latitude bands, leading to a minimum value $f_{min}\simeq 0.035$ for $E\geq 53$~EeV and $b_\text{SG}(\text{max})=19^\circ$, in which case 89 events are observed while 69.7 would be expected on average in isotropic simulations. The penalised probability for obtaining a smaller value of $f_{min}$ in isotropic simulations after a similar scan is $P=22\%$.

 The corresponding results for circular windows around the Galactic Center are shown in Figure~\ref{f.gc}. The minimum $f_\text{min}$ is obtained  for an angular radius around the GC of $19^\circ$ and for $E\geq 60$~EeV, for which 12 events are observed while 5.6 are expected on average for isotropic simulations. The penalised probability in this case is $P\simeq 29$\%, so that no significant excess results for any of the cases considered in this section.

\section{Search for Cross-Correlations with Astrophysical Catalogs}

In this section we search for correlations of the set of arrival directions with the celestial distribution of potential nearby cosmic ray sources. We choose approximately uniform and complete catalogs, namely the 2MRS catalog of galaxies \citep{2mrs}, the Swift-BAT \citep{Swift} X-ray catalog of AGNs\footnote{The 2MRS and Swift catalogs have been updated after our earlier study of correlations in \citet{app10,icrc11}.} and a catalog of radio galaxies with jets recently compiled in \citet{radiog}. 
 The three samples are quite complementary in identifying possible cosmic ray sources.
The normal  galaxies that dominate the 2MRS catalog may  trace the  locations of gamma ray bursts and/or  fast spinning newborn pulsars, whereas X-rays observed by
Swift identify AGNs hosted mainly by spiral galaxies, and the radio emission catalog selects extended jets and radio lobes of AGNs hosted mainly by elliptical galaxies.

The {2MASS Redshift Survey (2MRS)} \citep{2mrs} maps the distribution of galaxies in the nearby universe. It covers 91\% of the sky, except for Galactic latitudes  $|b|<5^\circ$ (and $|b|<8^\circ$ for longitudes within 30$^\circ$ of the Galactic Center). In the region covered it is essentially complete (at 97.6\%) for magnitudes brighter than $K_s=11.75$. It contains 43,533 galaxies with measured redshift\footnote{We adopt hereafter a Hubble constant of $H_0=70$~km/s/Mpc and the effective distances considered are taken as $D\equiv z c/H_0$, with $z$ the source redshift obtained from the catalog.}: 37,209 of them are within 200~Mpc and 16,422 within 100~Mpc. About 90\% of its objects have a redshift $z<0.05$, which is the range of distances of interest for UHECR correlation studies due to the effects of the GZK horizon.

The {Swift-BAT 70-month X-ray catalog} \citep{Swift} includes sources detected in 70 months of observation of the BAT hard X-ray detector on the Swift gamma-ray burst observatory. It contains a total of 1210 objects: 705 of them are AGN-like (Seyfert I and II, other AGNs, blazars and QSOs) with measured redshift. The catalog is complete over 90\% of the sky for fluxes  $>13.4{\times} 10^{-12}$ erg/(s cm$^2$), measured  in the X-ray band from 14 to 195 keV (note that the completeness of the subsample of AGNs with measured redshifts maybe slightly different). We use this cut in flux hereafter to have a more uniform sample of nearby AGNs. 489 AGN-like objects survive the cut: 296 of them are within 200~Mpc and 160 are within 100 Mpc.

The third catalog we use is a {compilation of radio galaxies} produced in \citet{radiog}. This is a combination of catalogs of observations at 1.4~GHz (NRAO VLA Sky Survey \citep{NVSS}) and 843~MHz (Sydney University Molonglo Sky Survey \citep{SUMSS}), with redshifts of associated objects taken from 2MRS. A flux limit of 213~mJy (289~mJy) at 1.4~GHz (843~MHz) is imposed to the objects from each respective catalog, which would correspond to the flux of Cen~A as seen from a distance of about 200~Mpc. We select from this catalog the radio galaxies having jets, which constitute a set of attractive candidates for UHECR sources. There are in total 407 such jetted radio galaxies: 205 are within 200 Mpc and 56 are within 100 Mpc (for this catalog we compute the distance using the redshift corrected for peculiar velocities that are also provided). We note that the majority of these radio galaxies are different from the Swift-BAT AGNs detected in X-rays,  being their overlap of only about 5\%. It is  also important to keep in mind that although we analyze each catalog individually, 
it is possible that different types of sources (i.e., from different catalogs) might be contributing to the overall UHECR fluxes.

Below we first study the cross-correlation with the three flux-limited catalogs (with the flux limits just described), including objects up to different maximum distances. This selection is based on the apparent luminosity, and is motivated by the fact that nearby sources may contribute significantly to the fluxes (in their corresponding electromagnetic band as well as in CRs) even if they are intrinsically fainter than far away sources. 
In the case of the AGNs in the Swift and radio-galaxy catalogs we also scan on the measured intrinsic luminosity of the objects. This is motivated by the fact that the maximum CR energy $E_\text{max}$ achievable at the sources may be linked to the intrinsic electromagnetic bolometric luminosity ${\cal L}$ of the source. In particular one could expect that $(E_\text{max}/Z)^2\propto {\cal L}$ if the energy density in the magnetic field is in equipartition with the energy in synchrotron emitting electrons in the acceleration region (see, e.g., \citet{fa09}). Hence, it might happen that only sources intrinsically brighter than some given luminosity are able to accelerate CRs above the threshold energies considered in this paper.  On the other hand, for the radio galaxies the luminosity is also correlated with the Fanaroff-Riley class, with FRII galaxies being generally brighter than FRI ones.

\subsection{Cross-Correlation with Flux-Limited Samples}

The basis of the cross-correlation technique is a counting of the number of pairs between UHE events and objects in the chosen catalogs. In a similar way to the analyses described in previous sections, we scan over energy threshold ($40\ {\rm EeV}\leq E_\text{th}\leq80$~EeV) and over the angular scale ($1^\circ\leq\psi\leq30^\circ$). We also consider different maximum distances to the objects, $D$, scanning on this from 10~Mpc up to 200~Mpc, in steps of 10~Mpc. To find excesses of pairs we compare their observed number with that resulting from isotropic simulations. For each considered distance $D$ we first calculate the fraction of isotropic simulations having an equal number to or more pairs than the data, $f(\psi,E_\text{th})$, and then we look for its minimum, $f_\text{min}$. The post-trial probability, $P$, is calculated as the fraction of isotropic simulations which, under similar scans over $E_\text{th}$ and $\psi$ for each considered $D$, lead to a value of $f_\text{min}$ smaller than the one obtained with the data. 

\begin{figure}[h!]
\centerline{
\includegraphics[height=2.in,angle=-0]{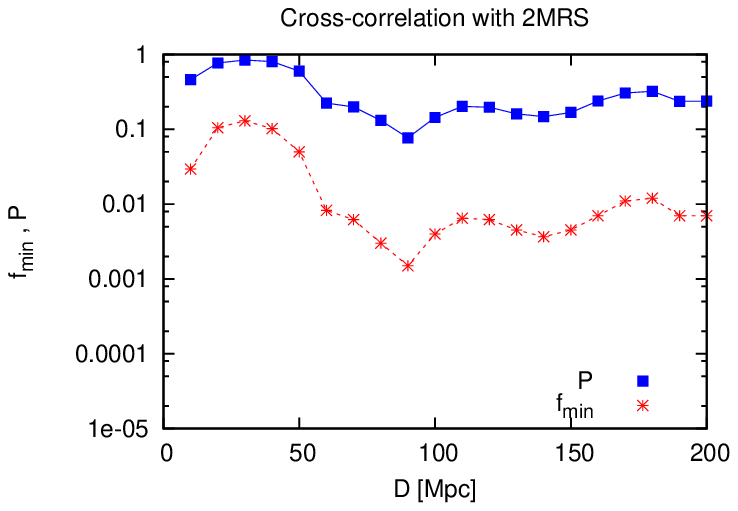}
\includegraphics[height=2.in,angle=-0]{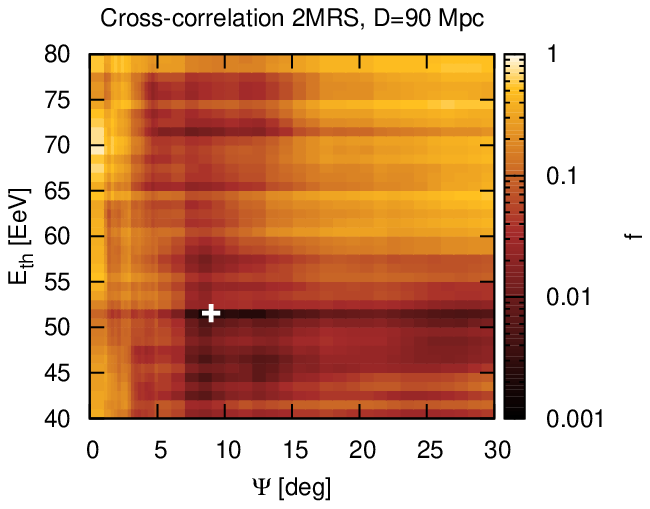}}
\centerline{\includegraphics[height=2.in,angle=-0]{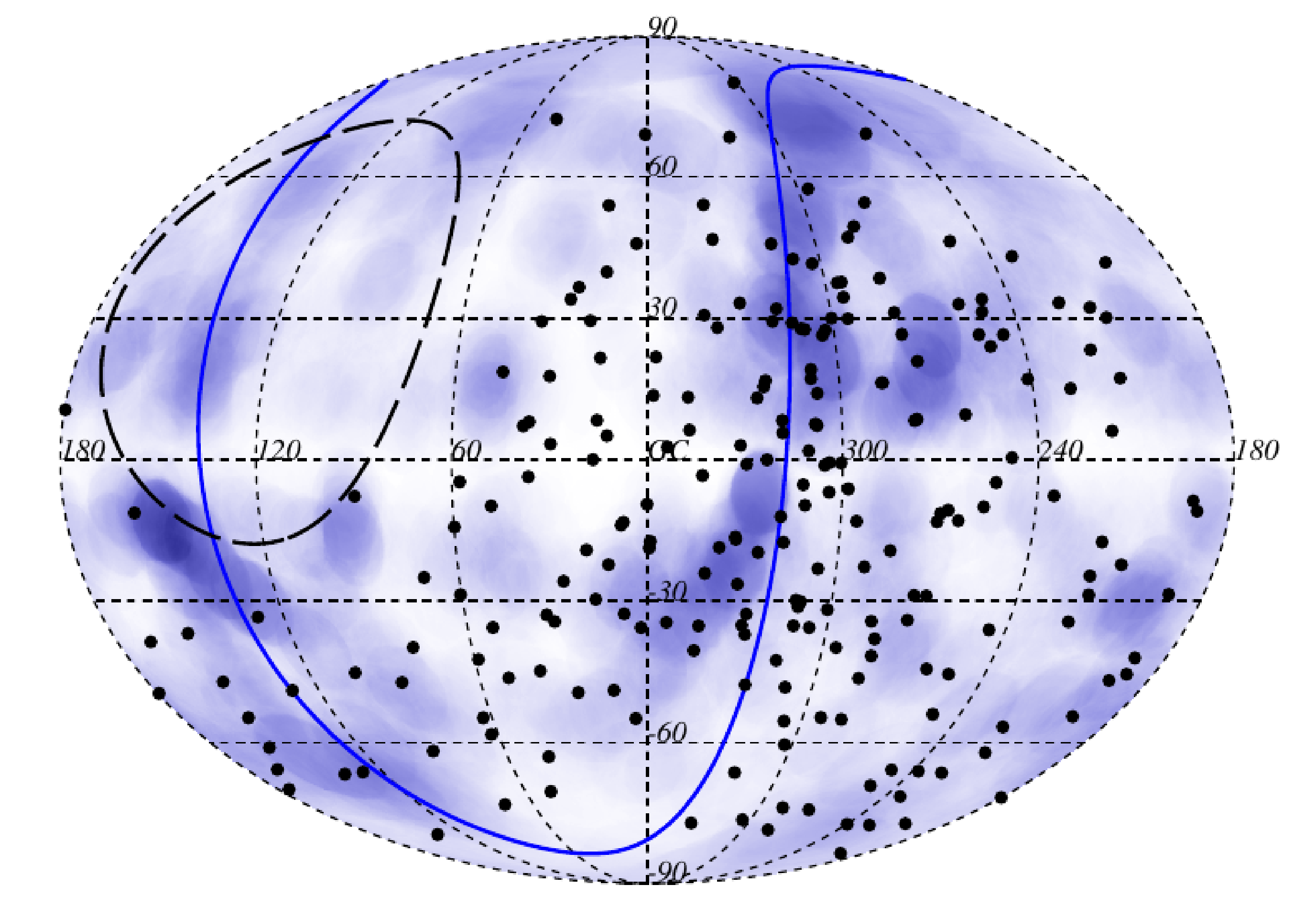}}
\caption{Cross-correlation of events with the galaxies in the 2MRS catalog. The top-left panel shows the values of $f_\text{min}$ and $P$ as a function of the maximum distance $D$ to the galaxies considered. The top-right panel shows the results of the scan in $\psi$ and $E_\text{th}$ for the value $D=90$~Mpc corresponding to the minimum values in the top-left plot. The bottom plot shows the sky distribution (in Galactic coordinates) of the events with $E\geq 52$~EeV (black dots). Blue fuzzy circles of $9^\circ$ radius are drawn around all of the 2MRS objects closer than 90~Mpc. The dashed line is the field-of-view limit for the Auger Observatory (for $\theta\leq 80^\circ$) and the blue solid line corresponds to the Super-Galactic Plane.} 
\label{f.2mrs}
\end{figure}

Figure~\ref{f.2mrs} displays the results for the case of the 2MRS catalog. The top-left panel shows $f_\text{min}$ (asterisks) and $P$ (squares) obtained for each distance $D$. 
The minimum values are observed for $D=90$~Mpc, for which $f_\text{min}\simeq 1.5{\times} 10^{-3}$ and $P\simeq 8$\%. The top-right panel in the figure shows the distribution of $f(\psi,E_\text{th})$ as a function of energy threshold and angle for the value $D=90$~Mpc giving rise to the minimum probability. The local minimum (indicated with a cross) is observed for $\psi=9^\circ$ and $E_\text{th}=52$~EeV.
For these values, 20,042 pairs are expected on average from isotropic realizations while 23,101 are observed in the data. Considering the penalization due to the scan in $D$ performed, the probability ${\cal P}$ to obtain a value of $P$ smaller than 8\% from isotropic distributions for any value of $D$ is ${\cal P}\simeq 24$\%.
Finally, the bottom panel of the figure displays the map of the events with $E\geq 52$~EeV (black dots).
Also drawn is a blue fuzzy circle around each 2MRS galaxy closer than 90~Mpc.
 All of those circles have radius $9^\circ$, which is the value for which
 the cross-correlation has maximum significance. Given the very large number of galaxies in 2MRS, essentially all events  are within 9$^\circ$ of at least one galaxy. Events falling in regions of the plot with denser color will have more galaxies within 9$^\circ$ and hence will contribute more pairs to the cross-correlation at this angular scale.

\begin{figure}[h!]
\centerline{\includegraphics[height=2.in,angle=-0]{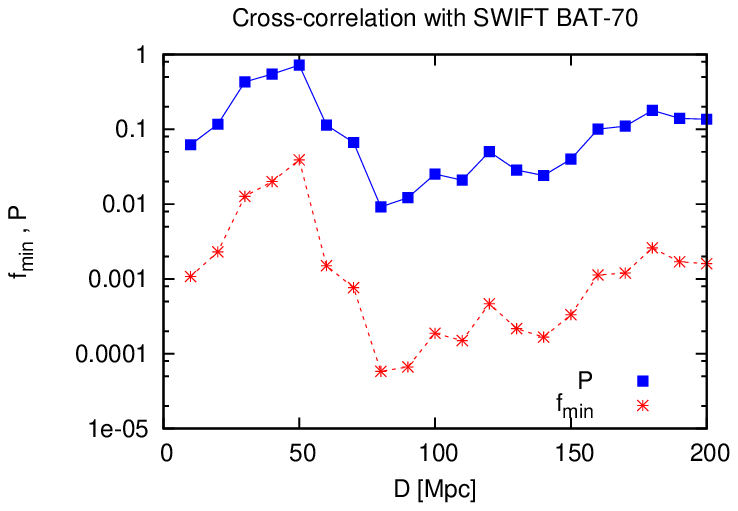}
\includegraphics[height=2.in,angle=-0]{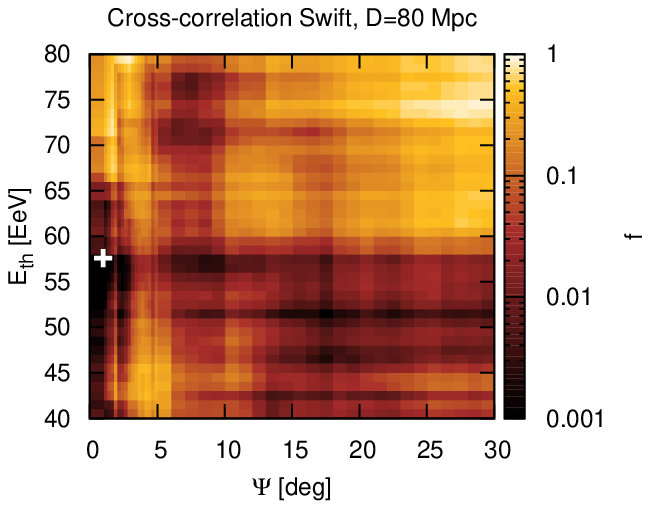}}
\bigskip
\centerline{\includegraphics[height=2.in,angle=-0]{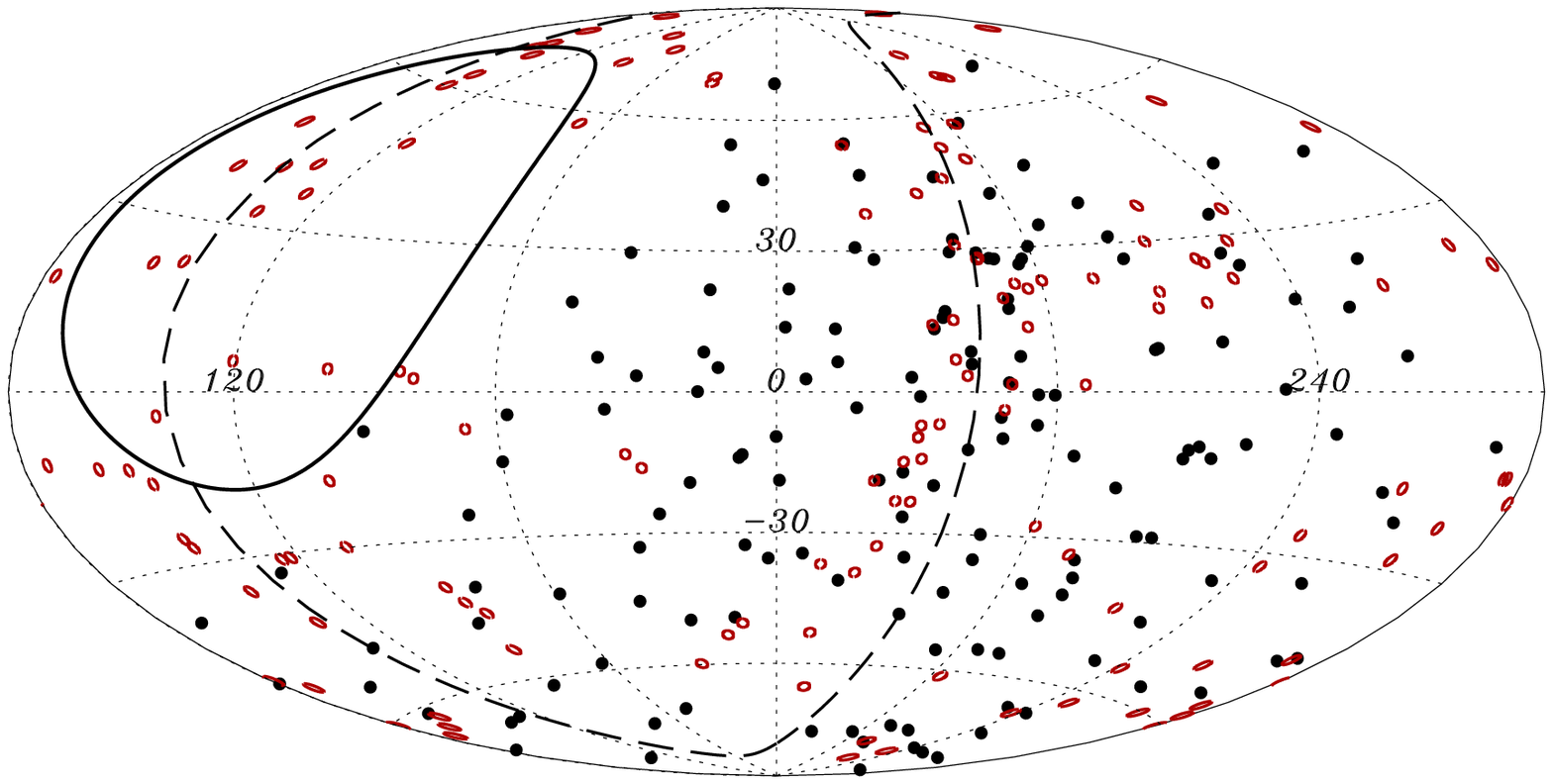}}
\caption{Cross-correlation of events with the AGNs in the Swift-BAT catalog. The top-left panel shows the values of $f_\text{min}$ and $P$ as a function of the maximum distance $D$ to the AGNs considered. The top-right panel shows the results of the scan in $\psi$ and $E_\text{th}$ for the value $D=80$~Mpc corresponding to the minimum values in the top-left plot. The bottom plot shows the sky distribution (in Galactic coordinates) of the events with $E\geq 58$~EeV (black dots). Red circles of 1$^\circ$ radius are drawn around the BAT AGNs closer than 80~Mpc. } 
\label{f.Swift}
\end{figure}

Similar plots to those presented above are included in Figure~\ref{f.Swift} for the case of the Swift-BAT catalog. As shown in the top-left panel of the figure, the minimum values are reached here for $D=80$~Mpc, where $f_\text{min}\simeq 6{\times} 10^{-5}$ and $P\simeq 1$\%. Correspondingly, the top-right panel in the figure shows $f(\psi,E_\text{th})$ as a function of energy and angle at $D=80$~Mpc. 
The local minimum (indicated with a cross) is at $\psi=1^\circ$ and $E_\text{th}=58$~EeV, where 9 pairs are observed and 1.6 are expected on average. 
After accounting for the penalization due to the scan performed in $D$, the probability of obtaining a value of $P$ smaller than 1\% from isotropic distributions for any value of $D$ is ${\cal P}\simeq 6$\%. Finally, we show  the map of events and objects in the bottom panel. Given the minimum found, we include events with $E\geq 58$~EeV and draw circles of 1$^\circ$ radius around the BAT AGNs closer than 80~Mpc. 

\begin{figure}[h!]
\centerline{\includegraphics[height=2.in,angle=-0]{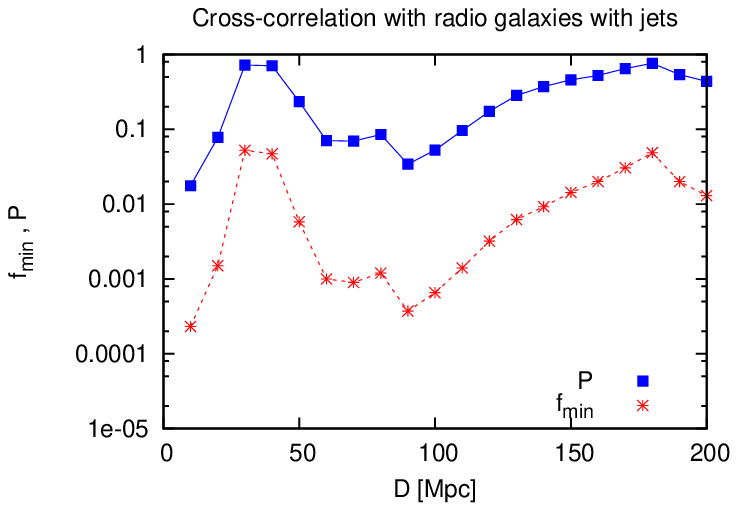}
\includegraphics[height=2.in,angle=-0]{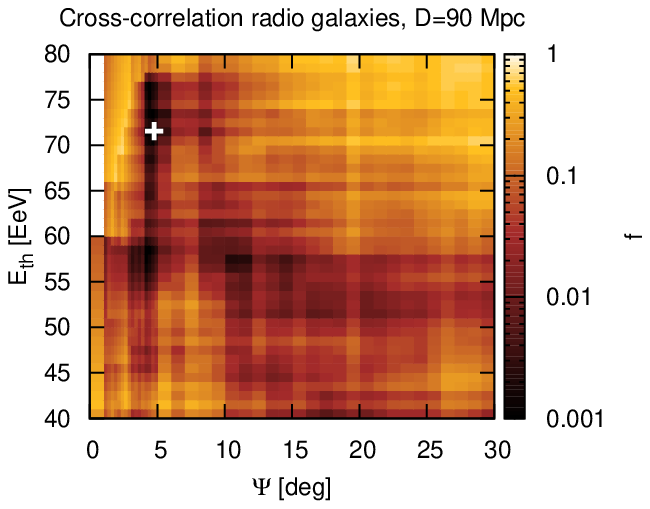}}
\bigskip
\centerline{\includegraphics[height=2.in,angle=-0]{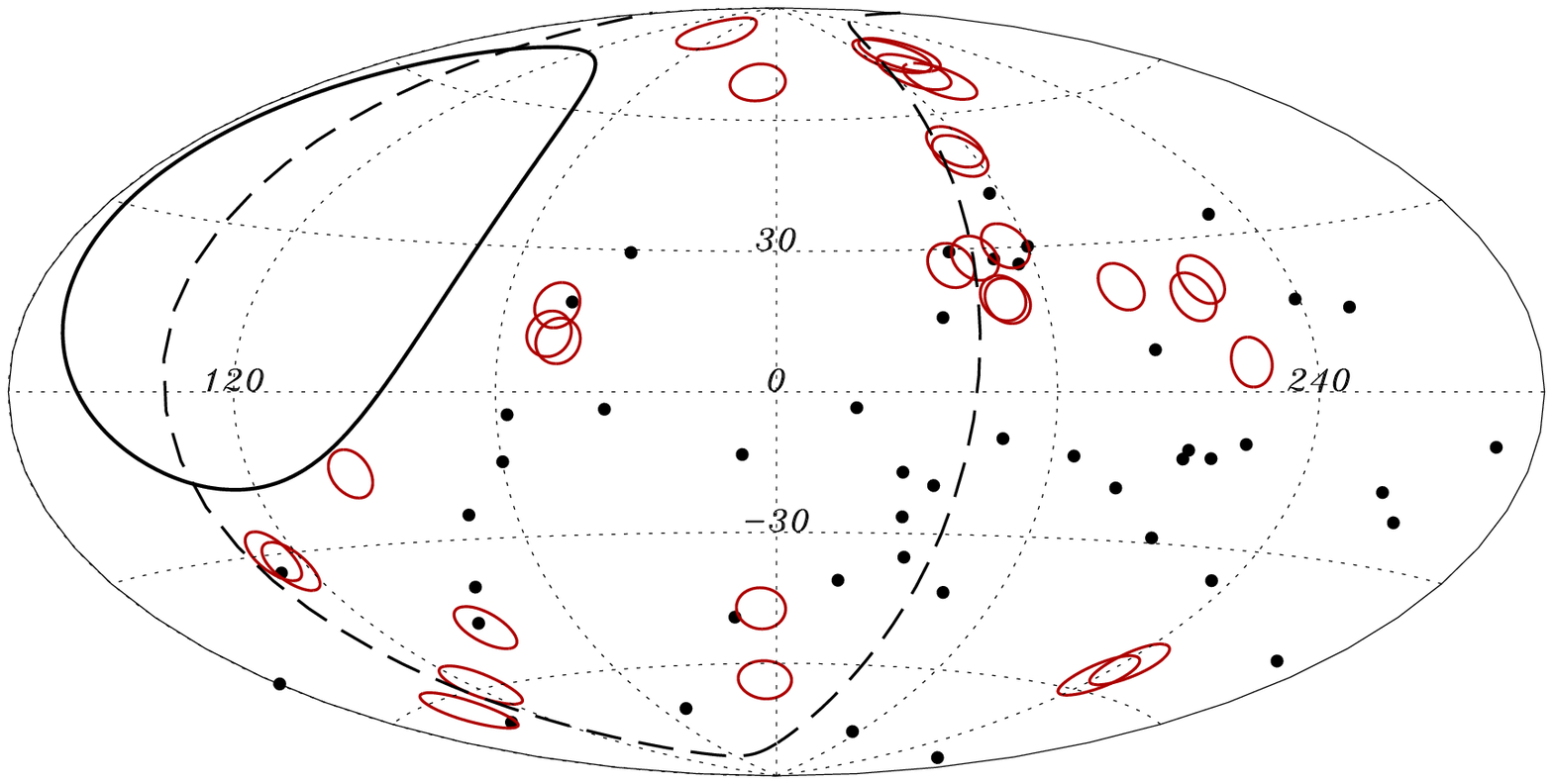}}
\caption{Cross-correlation of events with the AGNs in the catalog of radio galaxies with jets. The top-left panel shows the values of $f_\text{min}$ and $P$ as a function of the maximum distance $D$ to the AGNs considered. The top-right panel shows the results of the scan in $\psi$ and $E_\text{th}$ for the value $D=90$~Mpc corresponding to the (second) minimum in the top-left plot. The bottom plot shows the sky distribution (in Galactic coordinates) of the events with $E\geq 72$~EeV (black dots). Red circles of 4.75$^\circ$ radius are drawn around the radio galaxies closer than 90~Mpc.} 
\label{f.radiog}
\end{figure}

The results of the cross-correlation with jetted radio galaxies are shown in Figure~\ref{f.radiog}. The minimum value $f_\text{min}\simeq 2{\times} 10^{-4}$, with $P\simeq 1.4$\%, is obtained for $D=10$~Mpc (see top-left panel). The only object included in this catalog within such a distance is the Centaurus~A galaxy. Since the correlation with Cen~A is discussed separately in the next section, we consider here the second minimum, which is found for $D=90$~Mpc. This minimum corresponds to $f_\text{min}\simeq 4{\times} 10^{-4}$ and $P\simeq 3.4$\%. The top-right panel in the figure thus shows the results of the scan in angle and energy for $D=90$~Mpc. 
The minimum occurs for $\psi=4.75^\circ$ and $E_\text{th}=72$~EeV, where 13 pairs are observed in the data and 3.2 are expected on average. 
The chance probability for getting $P\leq 1.4$\% (corresponding to the absolute minimum found)  for any value of $D$  is ${\cal P}\simeq 8$\%. As was done for the other catalogs, the bottom panel displays the map of events and objects corresponding to the minimum found, i.e., $E\geq 72$~EeV and $D=90$~Mpc. Circles of radius 4.75$^\circ$ are drawn around every radio galaxy and the events are indicated with black dots.

While the cross-correlation analysis does not provide us with a significant indication of excesses of pairs with any of the catalogs considered, at any energy, distance and angle, we note that all of them yield minima  for similar maximum distances to the objects (${\sim}80$ to 90~Mpc) although for different threshold energies and angular scales. 
 The fact that the distances are similar for the three catalogs is actually expected given the existing correlations between catalogs, since AGNs are preferentially located in regions of high density of galaxies. On the other hand, the preference towards $D\simeq 80$~Mpc is mostly due to the fact that for this value the whole Centaurus Supercluster gets included and in this region there is an excess of high-energy events.

\subsection{Cross-Correlation with Bright AGNs}

We present here the results of a scan over the minimum source luminosities, considering for the Swift AGNs the reported X-ray band luminosity $\mathcal{L}_\text{X}$  and  for the radio-galaxy sample the reported radio luminosity $\mathcal{L}_\text{R}$, computed  per logarithmic energy bin at 1.1~GHz. For Swift we scan from $\mathcal{L}_\text{X}=10^{42}$~erg/s up to $10^{44}$~erg/s, while for the radio galaxies we scan from  $\mathcal{L}_\text{R}=10^{39}$~erg/s up to $10^{41}$~erg/s, considering three logarithmic steps per decade, for a total of 7 luminosity values in each case. These luminosity values cover most of the range spanned by the actual luminosities of the AGNs that are present in the catalogs (just 10 AGNs from the Swift sample have $\mathcal{L}_\text{X}<10^{42}$~erg/s, while only 3 AGNs from the radio-galaxy sample have $\mathcal{L}_\text{R}<10^{39}$~erg/s). Given the additional scan performed in $\mathcal{L}$, we do a slightly coarser scan in $D$, using 20~Mpc steps to cover from 10~Mpc up to 190~Mpc.

\begin{figure}[h!]
\centerline{\includegraphics[height=2.in,angle=-0]{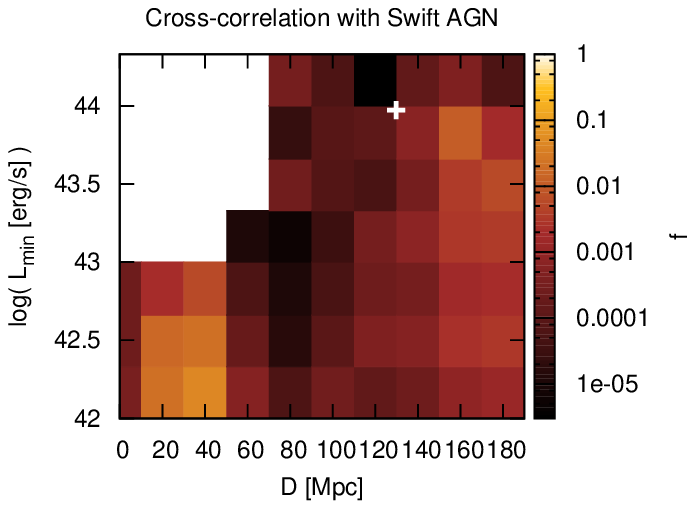}
\includegraphics[height=2.in,angle=-0]{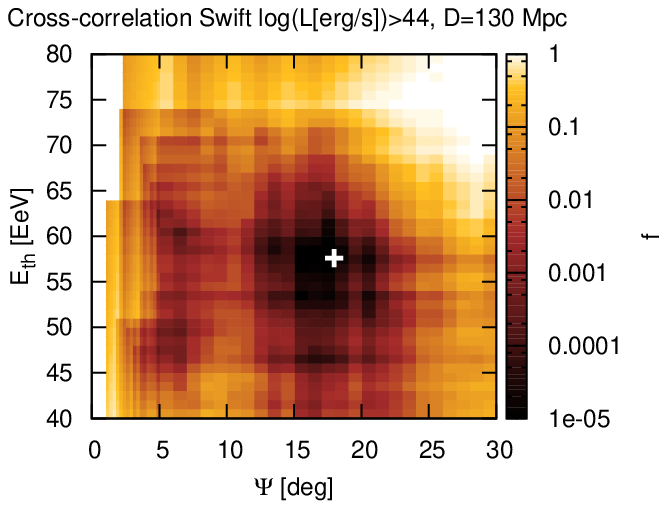}}
\centerline{\includegraphics[height=2.in,angle=-0]{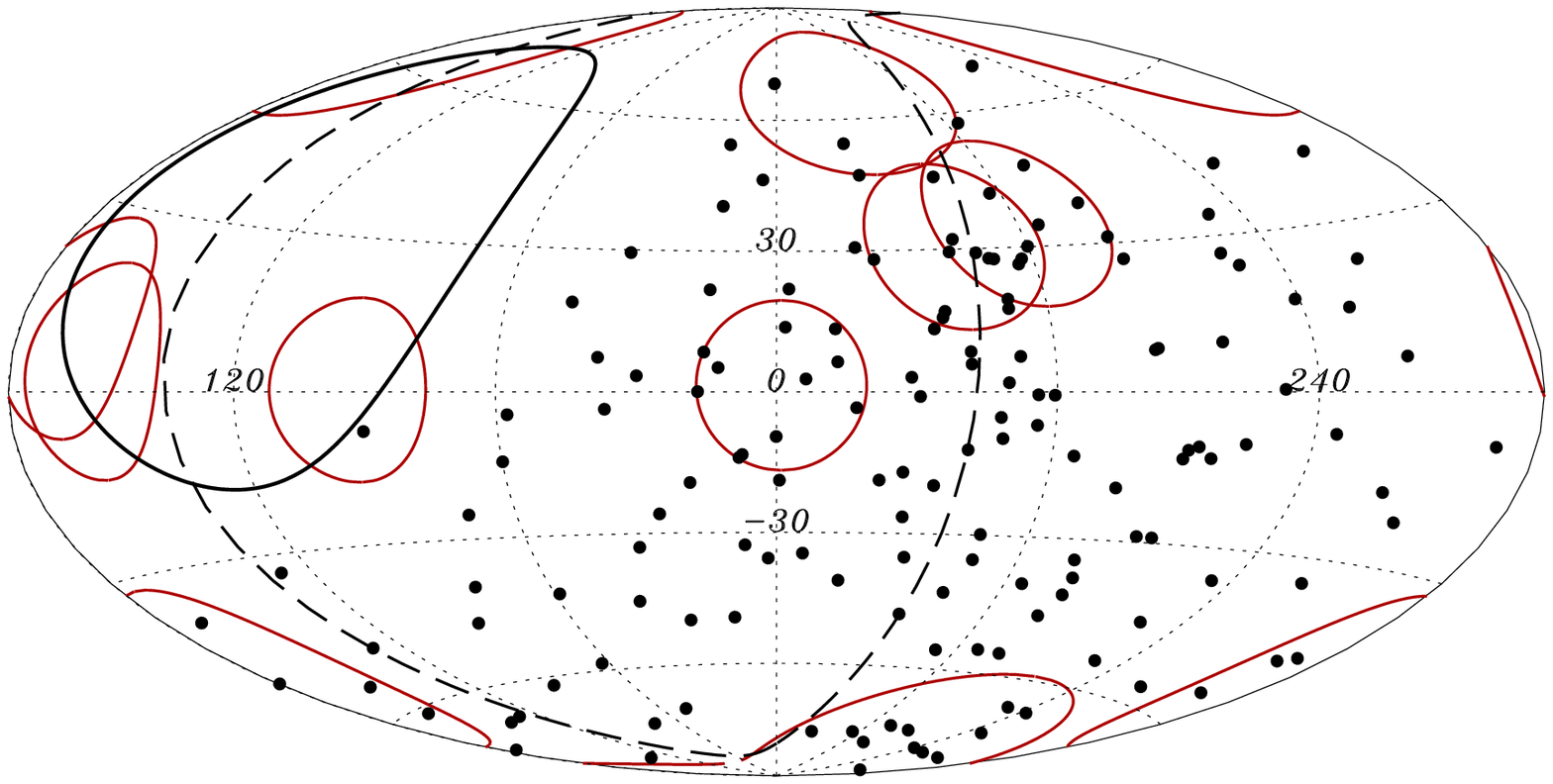}}
\caption{Cross-correlation of events with the AGNs in the Swift catalog as a function of $D$ and ${\cal L}_\text{min}$ (top-left panel) and detail of the scan in $\Psi$ and $E_\text{th}$ for the minimum found (top-right panel). The bottom map (in Galactic coordinates) shows the events with $E\geq 58$~EeV together with the Swift AGNs brighter than $10^{44}$~erg/s and closer than 130~Mpc, indicated with red circles of $18^\circ$ radius.}
\label{f.swiftl}
\end{figure}

 Considering first  the Swift catalog,
we show in the top-left panel of Figure~\ref{f.swiftl}  the resulting values of $f_\text{min}$  as a function of the maximum AGN distance and the minimum adopted luminosity ${\cal L}_\text{min}$ in the respective bands (the white region in the top-left corner of the plot is due to the absence of nearby  objects above those luminosity thresholds). The  values of $f_\text{min}$ are obtained after scanning on $\Psi$ and $E_\text{th}$ as in the previous subsection. The minimum value of $f_\text{min}=2{\times} 10^{-6}$ is obtained for $D=130$~Mpc and ${\cal L}>10^{44}$~erg/s. The top-right panel shows the details of the scan in $\Psi$ and $E_\text{th}$ for  $D=130$~Mpc and ${\cal L}>10^{44}$~erg/s. The minimum corresponds to the values $\Psi=18^\circ$ and $E_\text{th}=58$~EeV. For these parameters there are 10~AGNs and 155~events, and 62~pairs are obtained between them while the isotropic expectation is 32.8. The probability to find values  $f_\text{min}<2{\times} 10^{-6}$ in isotropic simulations after making the same scan on $\Psi$, $E_\text{th}$, ${\cal L}_\text{min}$ and $D$ is ${\cal P}\simeq 1.3$\%.

The bottom plot in the figure is the map of events with $E\geq 58$~EeV (black dots) and the Swift AGN brighter than $10^{44}$~erg/s that are closer than 130~Mpc, represented in the map with red circles of 18$^\circ$ radius, which is the value of $\Psi$ found at the minimum. We see that the events that mostly contribute to the excess of pairs observed are those arriving from directions contained inside the circles centered on IC~4329A (at ($\ell,b)=(317.6^\circ,30.9^\circ))$, ESO 506-G027  (at  ($\ell,b)=(299.6^\circ,35.5^\circ)$),  AX J1737.4-2907 (at ($\ell,b)=(358.9^\circ,1.4^\circ)$), NGC 612  (at ($\ell,b)=(261.8^\circ,-77^\circ)$) and NGC 1142  (at ($\ell,b)=(175.9^\circ,-49.9^\circ)$)\footnote{One of the objects in the sample of 10~AGNs is the BLLac Mrk~421, a powerful gamma-ray emitter at ($\ell,b)=(179.9^\circ,65^\circ)$, which has been proposed as a candidate source for the hot spot observed by the Telescope Array \citep{ol14}. This object is in a low-exposure region near the border of the Auger field of view, and there are no events with $E > 58$~EeV within 18$^\circ$ of it.}.

\begin{figure}[h!]
\centerline{\includegraphics[height=2.in,angle=-0]{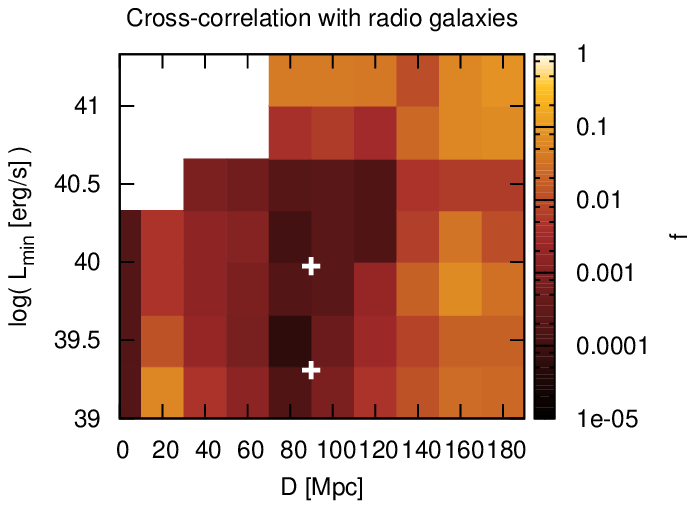}
\includegraphics[height=2.in,angle=-0]{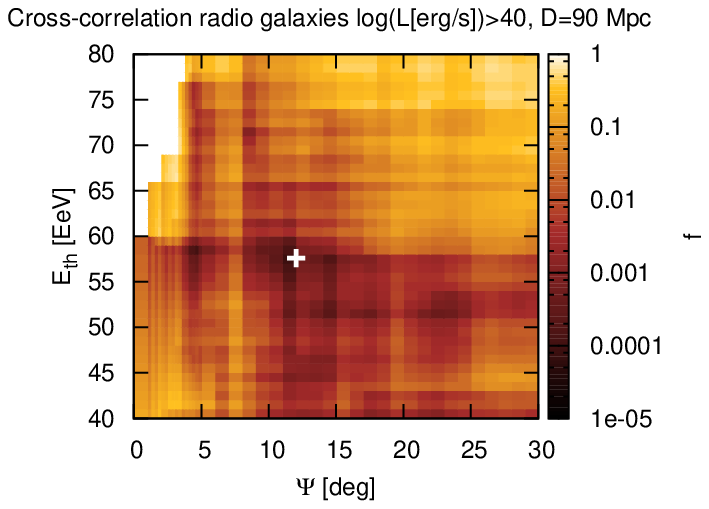}}
\centerline{\includegraphics[height=2.in,angle=-0]{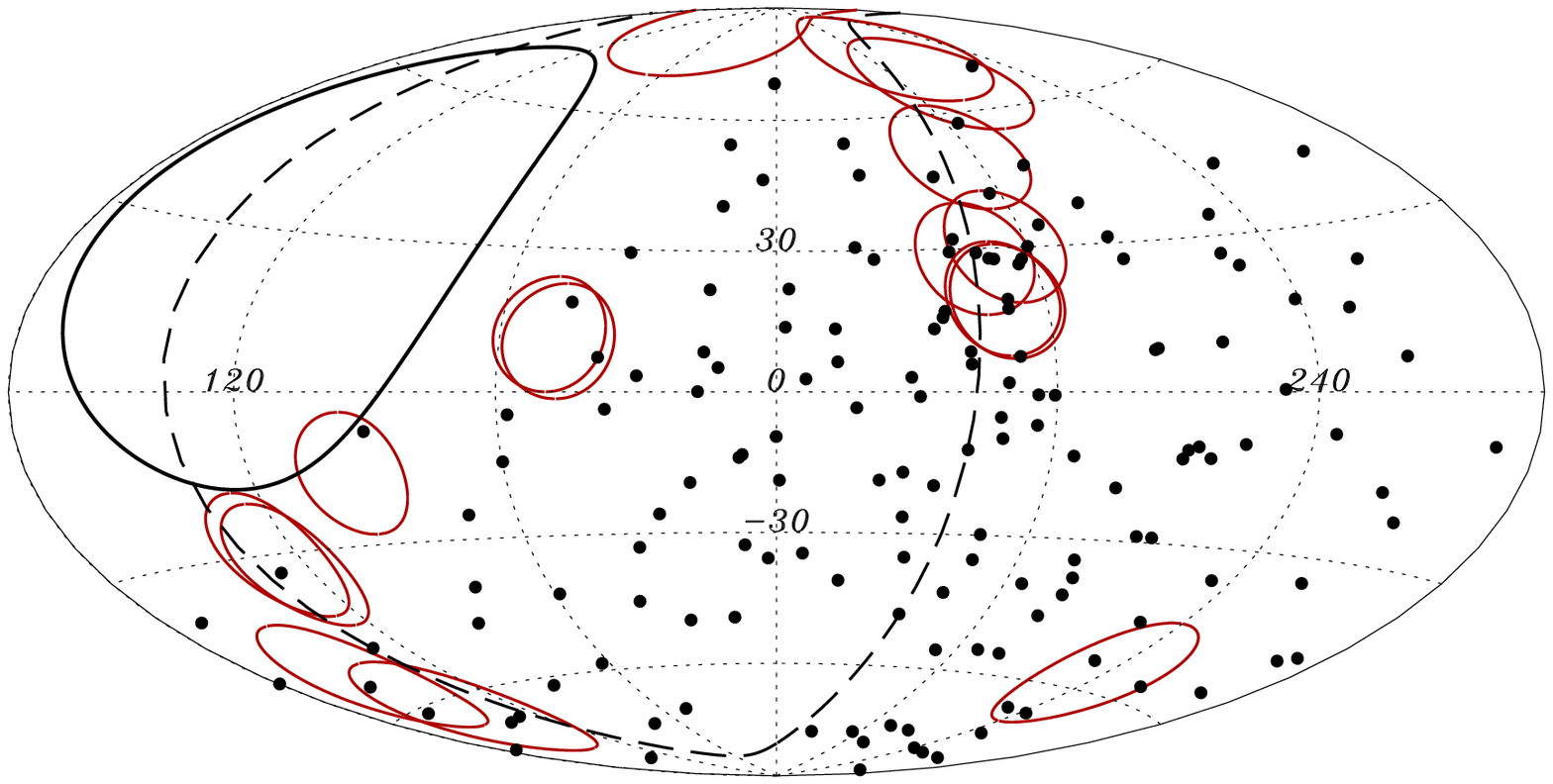}}
\caption{Cross-correlation of events with the radio galaxies as a function of $D$ and ${\cal L}_\text{min}$ (top-left panel) and detail of the scan in $\Psi$ and $E_\text{th}$ for  the second minimum found (top-right panel). The bottom map (in Galactic coordinates) shows the events with $E\geq 58$~EeV together with the radio galaxies brighter than $10^{40}$~erg/s and closer than 90~Mpc, indicated with red circles of $12^\circ$ radius (i.e., the parameters of the second minimum).}
\label{f.rgall}
\end{figure}

Figure~\ref{f.rgall} is similar but for the sample of radio galaxies. The scan in luminosity leads to two minima with very similar probabilities, both for $D=90$~Mpc (see the top-left panel). The first one has $f_\text{min}=5.1{\times} 10^{-5}$ and corresponds to ${\cal L}>10^{39.33}$~erg/s, $\Psi=4.75^\circ$ and $E_\text{th}=72$~EeV, the angle and energy being equal to the parameters already obtained in the previous subsection (Figure \ref{f.radiog}). The main difference is that 32~AGNs remain within 90~Mpc once the luminosity cut is imposed, compared to the original sample of 39~AGNs in the flux-limited sample, so that the expected number of pairs becomes 2.4 while 13 are actually observed. The second minimum has  $f_\text{min}=5.6{\times} 10^{-5}$ and corresponds to ${\cal L}>10^{40}$~erg/s. The top-right panel shows the scan in $\Psi$ and $E_\text{th}$ for this minimum, which leads to $\Psi=12^\circ$ and $E_\text{th}=58$~EeV. The bottom plot shows the map of the arrival directions of the events with $E\geq 58$~EeV (black dots)  and the radio galaxies within 90~Mpc, indicated with red circles of 12$^\circ$ radius. We see that most of the excess in the number of pairs arises from the events falling in the circles around the  radio galaxies in the Centaurus region. The globally penalized probability of getting $f_\text{min}<5.1{\times} 10^{-5}$ after a similar scan with the radio galaxies turns out to be in this case ${\cal P}\simeq 11$\%.

\section{The Centaurus A Region}

Centaurus~A is the nearest radio-loud active galaxy, at a distance of less than 4~Mpc. It is thus an obvious candidate source of UHECRs in the southern sky \citep{ro96}. In addition, the nearby Centaurus cluster is a large concentration of galaxies lying  in approximately the same direction and at a distance of ${\sim}50$~Mpc. The most significant localized excess of UHECR arrival directions reported earlier by the Pierre Auger Collaboration \citep{app10} was very close to the direction of Cen~A. In particular, we found 13 events with energy above 55~EeV in a circular window of radius $18^\circ$ centered on Cen~A, while 3.2 were expected in case of isotropy\footnote{We note however that the significance of the excess in this particular window of $18^\circ$ and for the rescaled energy threshold of 53~EeV did not grow with the additional data included in this work, for which $n_\text{obs}/n_\text{exp}=18/9$,  leading now to a cummulative binomial probability of $4\times 10^{-3}$.}. 
As shown in Section~4.1, the most significant excess observed in a blind search over the exposed sky with the present data set is also a region close to the direction of Cen~A. 

\begin{figure}[h!]
\centerline{\includegraphics[height=2.in,angle=-0]{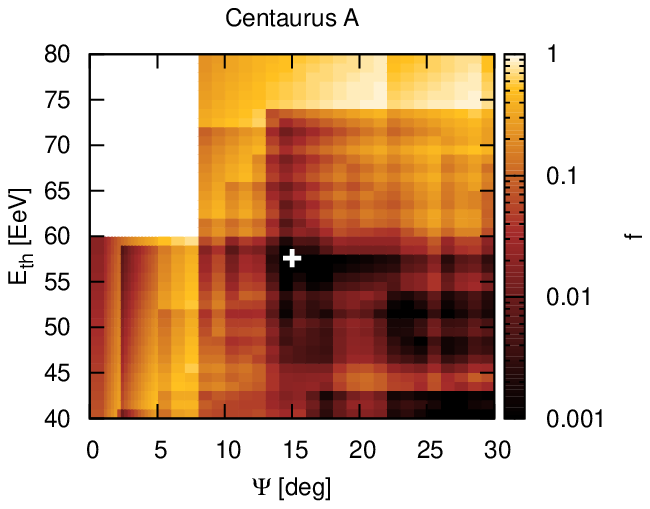} \ \ \   \ \  \includegraphics[height=1.8in,angle=-0]{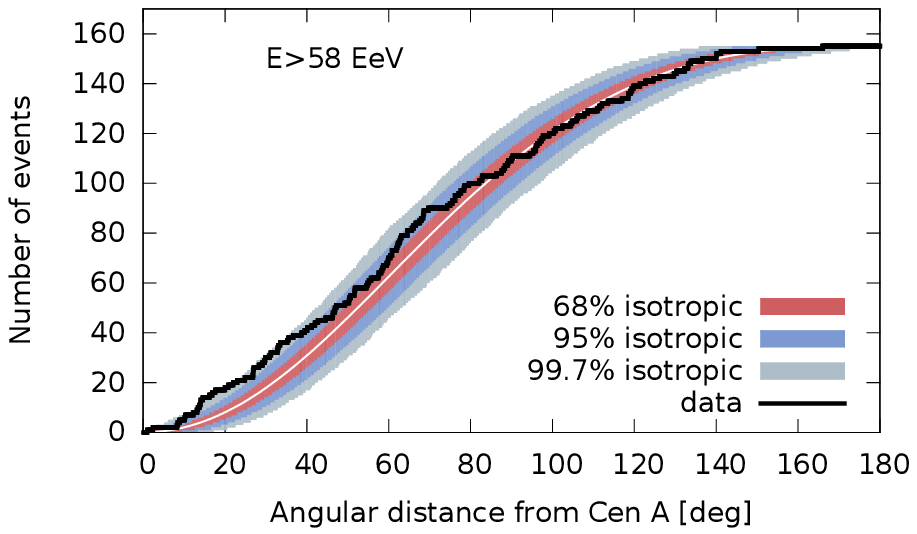}}
\bigskip
\centerline{\includegraphics[height=2.in,angle=-0]{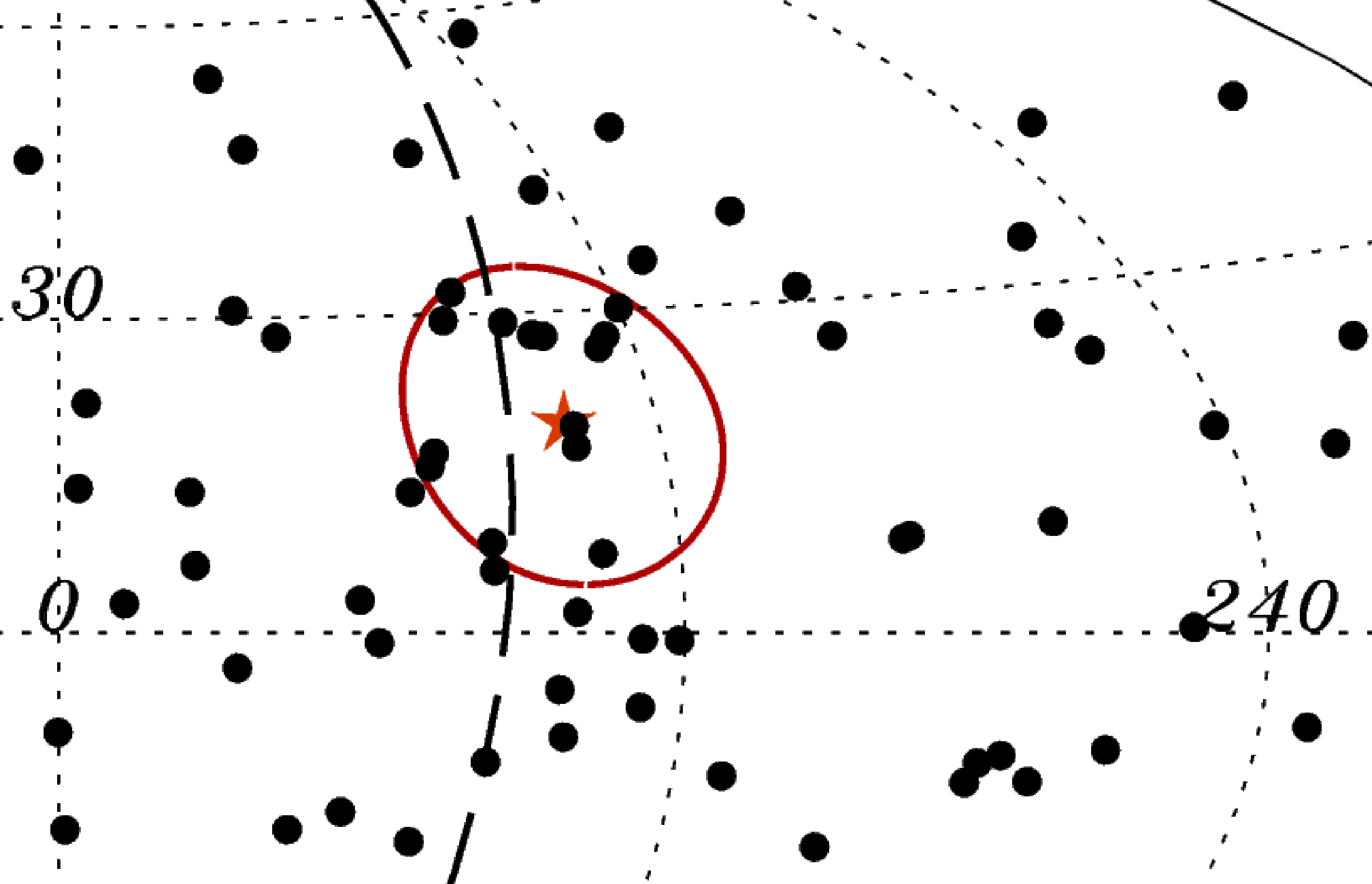}}
\caption{Correlation of events with the Cen~A radio galaxy as a function of the angular distance and the energy threshold, $E_\text{th}$ (top-left panel). The top-right panel shows the cumulative number of events  for the threshold $E_\text{th}=58$~EeV, exploring the whole angular range. The bottom panel displays the map (in Galactic coordinates) of the region around Centaurus~A, showing the arrival directions of the events with $E\geq 58$~EeV (black dots) and a red circle of 15$^\circ$ radius around the direction of Cen~A, indicated by a star.} 
\label{f.cena}
\end{figure}

In this section we search for cross-correlations of the arrival directions with the direction of Cen~A, $(\ell,b)=(-50.5^\circ,19.4^\circ)$. 
The search is performed by varying the energy threshold of events between 40~EeV and 80~EeV and by counting events in angular radii ranging from $1^\circ$ to $30^\circ$. To assess the significance of the observed number of events, we compare it to the one expected from isotropic simulations based on the same number of arrival directions as in the data. Figure~\ref{f.cena} (top-left panel) shows the fraction $f$ of those simulations that yield more than or an equal number of pairs to the data. The minimum value of $f$ is $f_\text{min}=2{\times} 10^{-4}$, corresponding to $E_\text{th}=58$~EeV and $\psi=15^\circ$.
There are 14 events (out of a total of 155) observed while 4.5 are expected on average from isotropic distributions. The fraction of isotropic simulated data sets that yield a smaller value of $f_\text{min}$ under a similar scan is ${\cal P}\simeq 1.4$\%. For completeness, we show in the top-right panel of the figure the number of events with energy above 58~EeV as a function  of the angular distance from Cen~A for the whole angular range, indicating also the 68, 95 and 99.7\% ranges obtained with isotropic simulations. The bottom panel displays the map in Galactic coordinates  of the Centaurus~A region, showing  the events with $E\geq 58$~EeV (black dots) and a  15$^\circ$ radius circle around the direction of Cen~A, indicated by a star.

\section{Discussion}

We have presented  several tests to search for signals of anisotropies in the arrival directions of the highest-energy events detected by the Pierre Auger Observatory from 2004 January 1 up to 2014 March 31. The main results obtained are summarized below.

We  first updated the fraction of events with energy above 53~EeV correlating with AGNs in the VCV catalog, obtaining a value of $28.1^{+3.8}_{-3.6}$\%, to be compared with 21\% for the isotropic expectation. This test then does not yield significant evidence of anisotropies above this particular energy threshold. Consequently, in all other exploratory analyses performed we have considered the data set down to an energy of 40~EeV.

A thorough search for overdense circular regions all over the sky and for different threshold energies led to the largest deviation from isotropy in a 12$^\circ$ radius window  centered at $(\alpha,\delta)=(198^\circ,-25^\circ$) and for events with energies above 54~EeV, but more significant excesses are obtained in 69\% of isotropic simulations under a similar scan. 
The autocorrelation of the events was also found to be  compatible with the expectations from an isotropic distribution. 

No significant excesses were found around the Galactic Center, the Galactic Plane, or the Super-Galactic Plane. 
This suggests that, if the deflections are not too large, at these energies the sources are unlikely to be Galactic and also that a non-negligible fraction of the flux arises from extragalactic sources that are not very close to the Super-Galactic Plane. 

The high degree of isotropy observed in all these tests of the distribution of UHECRs is indeed quite remarkable, certainly challenging original expectations that assumed only few cosmic ray sources with a light composition at the highest energies. If the actual source distribution were anisotropic, these results could be understood for instance as due to the large deflections caused by  the intervening magnetic fields if a large fraction of the CRs in this energy range were heavy, as is indeed suggested by mass-composition studies \citep{Augermass1,Augermass2}. Alternatively, it could also be explained in a scenario in which the number of individual sources contributing to the CR fluxes is large. Indeed, the lack of autocorrelation has been used in \citet{jcap12} to set lower bounds on the density of sources if the deflections involved are not large.

We have also studied the cross-correlation between events and nearby extragalactic objects in different flux-limited catalogs with the aim of identifying possible scenarios of UHECR sources. The parameters corresponding to the minima obtained when scanning in energy, distance and angular scale are listed in Table~1 (first three rows). The penalized probabilities that these minima are due to fluctuations of an isotropic background are of the order of a few percent. In all three cases the object distance corresponding to the minima is $D\simeq 80$ to 90~Mpc, although it happens for different angular scales and energy thresholds. When a further scan is performed on the minimum intrinsic AGN luminosity, additional minima  appear (see rows 4 and 5 in Table 1). We note that the penalized probability is ${\sim}1.3$\% for Swift AGNs within 130~Mpc and brighter than $10^{44}$~erg/s, corresponding to an excess of pairs for events above 58~EeV on angular scales of $18^\circ$, while for the radio galaxies the penalized probability is ${\sim}11$\%.

\begin{table}[hb]
\centerline{\begin{tabular}{c c c c c c c }
\hline
Objects & $E_\text{th}$  & $\Psi$  &  $D$ & ${\cal L}_\text{min}$  & $f_\text{min}$ & ${\cal P}$ \\
 &  [EeV] &  [$^\circ$] & [Mpc] & [erg/s] & &\\
\hline
2MRS Galaxies & 52 &  9 &  90 & - &  $1.5{\times} 10^{-3}$ &  24\% \\
Swift AGNs & 58 &  1 &  80 & - &  $6{\times} 10^{-5}$ &  6\% \\
Radio galaxies & 72 &  4.75 &  90 & - &  $2{\times} 10^{-4}$ &  8\% \\
\hline
Swift AGNs & 58 &  18 &  130 & $10^{44}$ &  $2{\times} 10^{-6}$ &  1.3\% \\
Radio galaxies & 58 &  12 &  90 & $10^{39.33}$ &  $5.6{\times} 10^{-5}$ &  11\% \\
\hline
Centaurus~A & 58 &  15 & - & - &  $2{\times} 10^{-4}$ &  1.4\% \\
\hline
\end{tabular}}
\caption{Summary of the parameters of the minima found in the cross-correlation analyses.}
\end{table}

Finally, considering circular windows  around the direction of Cen~A, the most significant indication of anisotropy appears for events with $E\geq 58$~EeV and for an angular radius of $15^\circ$. After penalizing for the scan on the angle and energy threshold this has a $1.4$\% probability of arising by chance from an isotropic distribution. Clearly the events contributing to the excess around the direction of Cen~A also contribute to the signals found in the cross-correlation searches performed against the different catalogs, which in general have an excess of objects in directions close to that of Cen~A. 

Overall, none of the tests performed yields a statistically significant evidence of anisotropy in the distribution of UHECRs.  It will be in any case interesting to follow with future data the evolution of the excesses found in the cross-correlation studies, particularly from Cen~A and from the bright AGNs.

\section*{Acknowledgments}
The successful installation, commissioning, and operation of the Pierre Auger Observatory would not have been possible without the strong commitment and effort from the technical and administrative staff in Malarg\"{u}e. 

We are very grateful to the following agencies and organizations for financial support: 
Comisi\'{o}n Nacional de Energ\'{\i}a At\'{o}mica, Fundaci\'{o}n Antorchas, Gobierno de la Provincia de Mendoza, Municipalidad de Malarg\"{u}e, NDM Holdings and Valle Las Le\~{n}as, in gratitude for their continuing cooperation over land access, Argentina; the Australian Research Council; Conselho Nacional de Desenvolvimento Cient\'{\i}fico e Tecnol\'{o}gico (CNPq), Financiadora de Estudos e Projetos (FINEP), Funda\c{c}\~{a}o de Amparo \`{a} Pesquisa do Estado de Rio de Janeiro (FAPERJ), S\~{a}o Paulo Research Foundation (FAPESP) Grants No. 2012/51015-5, 2010/07359-6 and No. 1999/05404-3, Minist\'{e}rio de Ci\^{e}ncia e Tecnologia (MCT), Brazil; Grant No. MSMT-CR LG13007, No. 7AMB14AR005, No. CZ.1.05/2.1.00/03.0058 and the Czech Science Foundation Grant No. 14-17501S, Czech Republic;  Centre de Calcul IN2P3/CNRS, Centre National de la Recherche Scientifique (CNRS), Conseil R\'{e}gional Ile-de-France, D\'{e}partement Physique Nucl\'{e}aire et Corpusculaire (PNC-IN2P3/CNRS), D\'{e}partement Sciences de l'Univers (SDU-INSU/CNRS), Institut Lagrange de Paris (ILP) Grant No. LABEX ANR-10-LABX-63, within the Investissements d'Avenir Programme  Grant No. ANR-11-IDEX-0004-02, France; Bundesministerium f\"{u}r Bildung und Forschung (BMBF), Deutsche Forschungsgemeinschaft (DFG), Finanzministerium Baden-W\"{u}rttemberg, Helmholtz Alliance for Astroparticle Physics (HAP), Helmholtz-Gemeinschaft Deutscher Forschungszentren (HGF), Ministerium f\"{u}r Wissenschaft und Forschung, Nordrhein Westfalen, Ministerium f\"{u}r Wissenschaft, Fors\-chung und Kunst, Baden-W\"{u}rttemberg, Germany; Istituto Nazionale di Fisica Nucleare (INFN), Ministero dell'Istruzione, dell'Universit\`{a} e della Ricerca (MIUR), Gran Sasso Center for Astroparticle Physics (CFA), CETEMPS Center of Excellence, Italy; Consejo Nacional de Ciencia y Tecnolog\'{\i}a (CONACYT), Mexico; Ministerie van Onderwijs, Cultuur en Wetenschap, Nederlandse Organisatie voor Wetenschappelijk Onderzoek (NWO), Stichting voor Fundamenteel Onderzoek der Materie (FOM), Netherlands; National Centre for Research and Development, Grants No. ERA-NET-ASPERA/01/11 and No. ERA-NET-ASPERA/02/11, National Science Centre, Grants No. 2013/08/M/ST9/00322, No. 2013/08/M/ST9/00728 and No. HARMONIA 5 - 2013/10/M/ST9/00062, Poland; Portuguese national funds and FEDER funds within Programa Operacional Factores de Competitividade through Funda\c{c}\~{a}o para a Ci\^{e}ncia e a Tecnologia (COMPETE), Portugal; Romanian Authority for Scientific Research ANCS, CNDI-UEFISCDI partnership projects Grants No. 20/2012 and No. 194/2012, Grants No. 1/ASPERA2/2012 ERA-NET, No. PN-II-RU-PD-2011-3-0145-17 and No. PN-II-RU-PD-2011-3-0062, the Minister of National  Education, Programme  Space Technology and Advanced Research (STAR), Grant No. 83/2013, Romania; Slovenian Research Agency, Slovenia; Comunidad de Madrid, FEDER funds, Ministerio de Educaci\'{o}n y Ciencia, Xunta de Galicia, European Community 7th Framework Program, Grant No. FP7-PEOPLE-2012-IEF-328826, Spain; Science and Technology Facilities Council, United Kingdom; Department of Energy, Contracts No. DE-AC02-07CH11359, No. DE-FR02-04ER41300, No. DE-FG02-99ER41107 and No. DE-SC0011689, National Science Foundation, Grant No. 0450696, The Grainger Foundation, USA; NAFOSTED, Vietnam; Marie Curie-IRSES/EPLANET, European Particle Physics Latin American Network, European Union 7th Framework Program, Grant No. PIRSES-2009-GA-246806; and UNESCO.

\newpage
\appendix
\section{LIST OF EVENTS}\label{AppendixA}
 
In this Appendix we give the arrival directions and energies of the 231 events\footnote{We note that out of the 69 events published in \citet{app10}, 5 turn out to have energies below 52~EeV with the present reconstruction and hence they do not appear in the table.} with $E\geq 52$~EeV and $\theta<80^\circ$ detected by the Pierre Auger Observatory from 2004 January 1 up to 2014 March 31. The threshold has been chosen so that it includes all of the events leading to the minimal probabilities in the cross-correlation studies performed with the different catalogs. The information about these events is collected in Table 2. The different columns are: year, Julian day for that year, zenith angle, energy, right ascension, declination, Galactic longitude and Galactic latitude.

\begin{figure}[h]
\begin{center}
\includegraphics[height=2.5in,angle=-0]{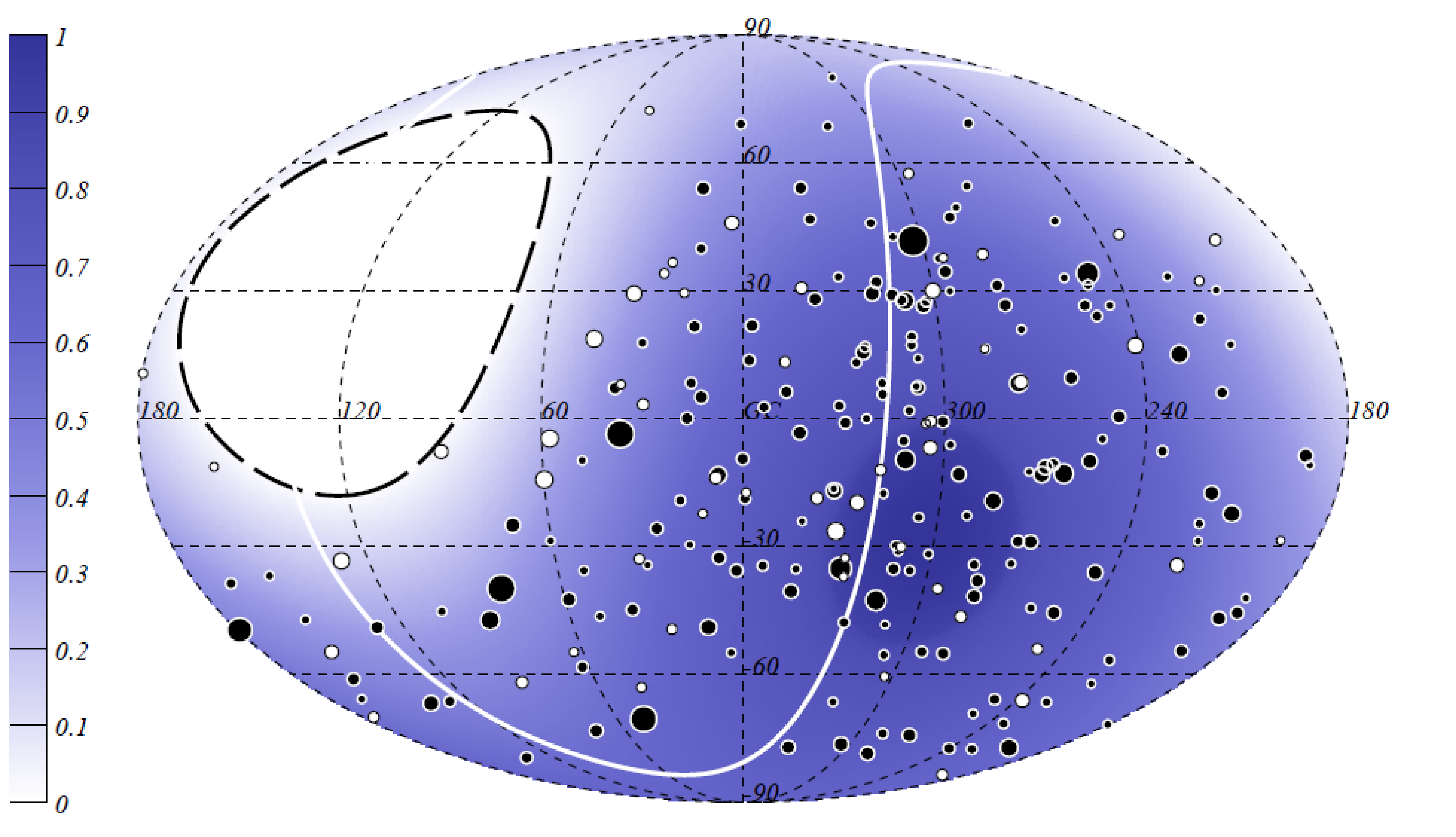}
\caption{Map in Galactic coordinates of the arrival directions of the events with $E\geq 52$~EeV. The black (white) circles correspond to {\it vertical} (\textit{inclined}) events. The size of each circle scales with the energy of the event. The color scale is proportional to the relative exposure.}
\label{f.mapevents}
\end{center}
\end{figure}

Figure \ref{f.mapevents} displays the arrival directions of these events  in Galactic coordinates. The dark filled circles correspond to the events in the vertical sample ($\theta\leq 60^\circ$) while the white filled circles correspond to those in the inclined sample ($60^\circ<\theta<80^\circ$).
The size of the circles scales with  the energy of the events. The background color in the map indicates the relative exposure of the Auger Observatory to different declinations. The white region is outside the field of view of the Auger Observatory for $\theta<80^\circ$.

\begin{deluxetable}{ c c r r r r r r}
\tabletypesize{\footnotesize}
\tablecolumns{8}
\tablewidth{0pt}
\tablecaption{List of the events with energies above 52~EeV  and $\theta<80^\circ$, the columns being: year, day, zenith angle $\theta$, energy $E$, right ascension $\alpha$, declination $\delta$ and Galactic longitude $\ell$ and latitude $b$.}
\tablehead{
Year & Julian & \multicolumn{1}{c}{ $\theta$} &  \multicolumn{1}{c}{$E$} & \multicolumn{1}{c}{$\alpha$}  & \multicolumn{1}{c}{$\delta$} &\multicolumn{1}{c}{ $\ell$} & \multicolumn{1}{c}{$b$} \\
     &  day &    \multicolumn{1}{c}{[$^\circ$]} &  [EeV] & \multicolumn{1}{c}{ [$^\circ$]} &    \multicolumn{1}{c}{[$^\circ$]} &  \multicolumn{1}{c}{[$^\circ$]} &  \multicolumn{1}{c}{[$^\circ$]} \\\
     }
     \startdata
2004 & 125 &   47.7 &   62.2 &  267.2 &  $-11.4$ &   15.5 &    8.4   \\
2004 & 142 &   59.2 &   84.7 &  199.7 &  $-34.9$ &  $-50.8$ &   27.7   \\
2004 & 177 &   71.5 &   54.6 &   12.7 &  $-56.6$ &  $-56.9$ &  $-60.5$   \\
2004 & 239 &   58.3 &   54.0 &   32.7 &  $-85.0$ &  $-59.1$ &  $-31.8$   \\
2004 & 282 &   26.3 &   58.6 &  208.1 &  $-60.1$ &  $-49.5$ &    1.9   \\
2004 & 339 &   44.6 &   78.2 &  268.4 &  $-61.0$ &  $-27.6$ &  $-16.9$   \\
2004 & 343 &   23.3 &   58.2 &  224.7 &  $-44.0$ &  $-34.1$ &   13.1   \\
2005 &  50 &   67.5 &   60.2 &   29.0 &  $-14.0$ &  174.9 &  $-70.0$   \\
2005 &  54 &   34.9 &   71.2 &   17.5 &  $-37.8$ &  $-76.0$ &  $-78.6$   \\
2005 &  63 &   54.4 &   71.9 &  331.2 &   $-1.3$ &   58.7 &  $-42.4$   \\
2005 &  81 &   17.1 &   52.1 &  199.1 &  $-48.5$ &  $-52.8$ &   14.1   \\
2005 & 186 &   57.5 &  108.2 &   45.6 &   $-1.7$ &  179.5 &  $-49.6$   \\
2005 & 233 &   65.4 &   61.9 &  278.4 &   $-1.3$ &   29.7 &    3.4   \\
2005 & 295 &   15.3 &   54.9 &  333.0 &  $-38.1$ &    4.4 &  $-55.0$   \\
2005 & 306 &   14.2 &   74.9 &  114.8 &  $-42.8$ & $-103.9$ &  $-10.0$   \\
2005 & 347 &   65.6 &   77.5 &   18.3 &   29.2 &  128.6 &  $-33.4$   \\
2006 &   5 &   30.9 &   78.2 &   18.9 &   $-4.7$ &  138.3 &  $-66.8$   \\
2006 &  35 &   30.8 &   72.2 &   53.6 &   $-7.8$ & $-165.9$ &  $-46.9$   \\
2006 &  55 &   37.9 &   52.8 &  267.6 &  $-60.6$ &  $-27.5$ &  $-16.4$   \\
2006 &  64 &   66.6 &   64.8 &  275.2 &  $-57.2$ &  $-22.6$ &  $-18.6$   \\
2006 &  81 &   34.0 &   69.5 &  201.1 &  $-55.3$ &  $-52.3$ &    7.3   \\
2006 & 100 &   33.7 &   54.7 &   28.8 &  $-16.4$ & $-179.9$ &  $-71.8$   \\
2006 & 118 &   57.3 &   56.3 &  322.5 &   $-2.0$ &   51.6 &  $-35.6$   \\
2006 & 126 &   65.2 &   82.0 &  299.0 &   19.4 &   57.6 &   $-4.7$   \\
2006 & 142 &   22.6 &   64.3 &  121.8 &  $-52.5$ &  $-93.0$ &  $-10.7$   \\
2006 & 160 &   76.5 &   60.7 &   52.7 &  $-43.4$ & $-109.6$ &  $-54.1$   \\
2006 & 185 &   58.8 &   89.0 &  349.9 &    9.3 &   88.4 &  $-47.3$   \\
2006 & 263 &   49.9 &   53.0 &   82.1 &   14.6 & $-169.9$ &  $-10.9$   \\
2006 & 284 &   54.5 &   54.0 &  142.3 &  $-13.1$ & $-114.3$ &   26.6   \\
2006 & 296 &   53.9 &   67.7 &   53.0 &   $-4.5$ & $-170.5$ &  $-45.6$   \\
2006 & 299 &   26.0 &   59.5 &  200.9 &  $-45.3$ &  $-51.2$ &   17.2   \\
2006 & 350 &   17.6 &   60.0 &  305.6 &  $-46.3$ &   $-6.4$ &  $-34.5$   \\
2007 &   9 &   54.0 &   53.8 &  321.0 &    8.1 &   60.4 &  $-28.7$   \\
2007 &  13 &   14.2 &  127.1 &  192.8 &  $-21.2$ &  $-57.1$ &   41.7   \\
2007 &  14 &   55.9 &   52.2 &  192.6 &   17.2 &  $-58.4$ &   80.1   \\
2007 &  69 &   30.4 &   60.0 &  200.2 &  $-43.4$ &  $-51.4$ &   19.2   \\
2007 &  84 &   17.5 &   60.8 &  143.4 &  $-18.1$ & $-109.4$ &   24.1   \\
2007 & 106 &   49.8 &   70.3 &   17.5 &   13.6 &  129.8 &  $-49.0$   \\
2007 & 145 &   24.0 &   68.4 &   47.5 &  $-12.8$ & $-164.0$ &  $-54.5$   \\
2007 & 161 &   41.9 &   53.6 &  137.3 &    6.2 & $-135.9$ &   33.4   \\
2007 & 166 &   79.6 &   54.9 &  245.8 &    8.5 &   22.9 &   36.7   \\
2007 & 186 &   44.9 &   61.5 &  219.5 &  $-53.9$ &  $-41.7$ &    5.8   \\
2007 & 193 &   17.9 &   79.7 &  325.5 &  $-33.4$ &   12.2 &  $-49.0$   \\
2007 & 203 &   55.3 &   57.0 &  265.9 &    5.9 &   30.5 &   17.8   \\
2007 & 205 &   76.5 &   61.9 &  195.5 &  $-63.4$ &  $-55.9$ &    $-0.6$   \\
2007 & 221 &   35.5 &   67.8 &  212.8 &   $-3.1$ &  $-21.6$ &   54.2   \\
2007 & 227 &   33.6 &   60.7 &  192.5 &  $-35.3$ &  $-57.3$ &   27.5   \\
2007 & 234 &   33.3 &   68.1 &  185.3 &  $-27.9$ &  $-65.2$ &   34.5   \\
2007 & 235 &   42.6 &   60.8 &  105.9 &  $-22.9$ & $-125.2$ &   $-7.7$   \\
2007 & 295 &   21.1 &   65.9 &  325.7 &  $-15.5$ &   37.8 &  $-44.8$   \\
2007 & 295 &   56.5 &   55.8 &   39.2 &   19.4 &  154.4 &  $-36.9$   \\
2007 & 314 &   76.7 &   52.5 &   59.6 &   38.3 &  158.5 &  $-11.3$   \\
2007 & 339 &   68.2 &   54.0 &  250.3 &    1.8 &   18.5 &   29.5   \\
2007 & 343 &   30.9 &   82.4 &   81.6 &   $-7.4$ & $-150.1$ &  $-22.3$   \\
2007 & 345 &   51.6 &   72.7 &  315.3 &  $-53.8$ &  $-16.0$ &  $-40.5$   \\
2008 &  10 &   77.1 &   80.2 &  271.1 &   19.0 &   45.2 &   18.7   \\
2008 &  13 &   16.8 &   64.2 &  252.7 &  $-22.7$ &   $-1.9$ &   13.7   \\
2008 &  18 &   50.2 &  111.8 &  352.6 &  $-20.8$ &   47.5 &  $-70.5$   \\
2008 &  36 &   28.3 &   65.3 &  187.5 &  $-63.5$ &  $-59.5$ &    $-0.7$   \\
2008 &  48 &   76.9 &   60.4 &   19.8 &  $-25.5$ & $-160.1$ &  $-83.6$   \\
2008 &  49 &   50.7 &   56.0 &   64.1 &  $-52.7$ &  $-98.5$ &  $-44.4$   \\
2008 &  51 &   20.7 &   53.3 &  202.0 &  $-54.9$ &  $-51.8$ &    7.6   \\
2008 &  52 &   31.7 &   56.2 &   82.8 &  $-15.8$ & $-141.2$ &  $-24.7$   \\
2008 &  72 &    4.4 &   52.4 &  184.4 &  $-32.4$ &  $-65.2$ &   30.0   \\
2008 &  87 &   38.9 &   73.1 &  220.6 &  $-42.8$ &  $-36.3$ &   15.5   \\
2008 & 118 &   36.2 &   62.9 &  110.2 &    $-0.9$ & $-142.9$ &    6.1   \\
2008 & 142 &   43.4 &   56.7 &  199.4 &    6.6 &  $-39.0$ &   68.5   \\
2008 & 184 &   53.7 &   55.7 &   33.0 &   11.0 &  152.8 &  $-47.2$   \\
2008 & 192 &   20.2 &   55.1 &  306.5 &  $-55.1$ &  $-17.1$ &  $-35.3$   \\
2008 & 205 &   53.1 &   56.7 &  358.9 &   15.5 &  103.6 &  $-45.2$   \\
2008 & 250 &   68.8 &   52.0 &   67.7 &    4.0 & $-168.7$ &  $-28.6$   \\
2008 & 264 &   44.4 &   89.3 &  116.0 &  $-50.6$ &  $-96.4$ &  $-12.9$   \\
2008 & 266 &   59.0 &   61.2 &  339.4 &  $-63.3$ &  $-35.4$ &  $-47.8$   \\
2008 & 268 &   49.8 &  118.3 &  287.7 &    1.5 &   36.5 &   $-3.6$   \\
2008 & 282 &   29.0 &   58.1 &  202.2 &  $-16.1$ &  $-44.2$ &   45.9   \\
2008 & 296 &   42.8 &   64.7 &   15.6 &  $-17.1$ &  137.9 &  $-79.6$   \\
2008 & 322 &   28.4 &   62.2 &   25.0 &  $-61.4$ &  $-67.1$ &  $-54.8$   \\
2008 & 328 &   47.2 &   63.1 &  126.4 &    5.3 & $-140.8$ &   23.4   \\
2008 & 329 &   47.9 &   66.9 &   28.9 &   $-2.7$ &  157.9 &  $-61.2$   \\
2008 & 331 &   50.7 &   52.6 &  304.4 &  $-26.2$ &   16.7 &  $-29.6$   \\
2008 & 337 &   30.8 &   65.8 &  275.2 &  $-14.4$ &   16.7 &     0.1   \\
2008 & 355 &   71.7 &   71.1 &  196.1 &  $-69.7$ &  $-55.9$ &   $-6.9$   \\
2008 & 362 &   31.5 &   74.0 &  209.6 &  $-31.3$ &  $-40.7$ &   29.4   \\
2009 &   7 &   59.2 &   61.0 &  286.3 &  $-37.8$ &    $-0.6$ &  $-18.$7   \\
2009 &  30 &   32.3 &   66.2 &  303.9 &  $-16.5$ &   26.8 &  $-25.8$   \\
2009 &  32 &   56.2 &   70.3 &     0.0 &  $-15.4$ &   75.0 &  $-73.2$   \\
2009 &  35 &   52.8 &   57.7 &  227.0 &  $-85.2$ &  $-54.2$ &  $-23.1$   \\
2009 &  39 &   42.4 &   64.1 &  147.2 &  $-18.3$ & $-106.5$ &   26.6   \\
2009 &  47 &   20.7 &   52.9 &   78.3 &  $-16.0$ & $-142.9$ &  $-28.8$   \\
2009 &  51 &    6.9 &   66.7 &  203.4 &  $-33.0$ &  $-47.0$ &   29.1   \\
2009 &  73 &   37.0 &   72.5 &  193.8 &  $-36.4$ &  $-56.2$ &   26.5   \\
2009 &  78 &   27.2 &   74.4 &  122.7 &  $-54.7$ &  $-90.7$ &  $-11.4$   \\
2009 &  78 &    8.2 &   59.0 &   26.7 &  $-29.1$ & $-134.5$ &  $-77.6$   \\
2009 &  80 &   18.4 &   65.8 &  251.4 &  $-35.8$ &  $-13.0$ &    6.3   \\
2009 &  80 &   44.4 &   63.8 &  170.1 &  $-27.4$ &  $-80.8$ &   31.3   \\
2009 &  83 &   68.6 &   56.2 &  249.1 &    9.1 &   25.3 &   34.1   \\
2009 & 140 &   27.2 &   55.1 &  330.8 &   $-8.9$ &   49.5 &  $-46.3$   \\
2009 & 160 &   40.9 &   52.8 &   43.9 &  $-25.4$ & $-143.4$ &  $-62.2$   \\
2009 & 162 &   78.2 &   70.5 &   39.4 &  $-34.5$ & $-122.6$ &  $-66.1$   \\
2009 & 163 &   41.2 &   71.9 &   23.3 &  $-40.2$ &  $-87.9$ &  $-74.3$   \\
2009 & 172 &    9.7 &   65.8 &  276.1 &  $-33.4$ &     0.1 &   $-9.4$   \\
2009 & 191 &   26.9 &   59.5 &  294.5 &  $-20.5$ &   19.1 &  $-19.2 $  \\
2009 & 197 &   51.7 &   52.2 &  129.4 &   15.2 & $-149.5$ &   30.2   \\
2009 & 202 &   60.8 &   63.6 &  358.2 &   $-2.8$ &   90.4 &  $-61.9$   \\
2009 & 212 &   52.7 &   55.3 &  122.5 &  $-78.5$ &  $-68.8$ &  $-22.8$   \\
2009 & 219 &   40.1 &   53.2 &   29.4 &   $-8.6$ &  166.2 &  $-65.8$   \\
2009 & 219 &   59.7 &   58.3 &  304.3 &  $-81.9$ &  $-48.3$ &  $-29.8$   \\
2009 & 237 &   78.4 &   70.0 &  325.8 &   42.8 &   90.1 &   $-7.8$   \\
2009 & 250 &   70.7 &   52.3 &  212.7 &   29.9 &   46.8 &   72.3   \\
2009 & 262 &   22.4 &   58.7 &   50.1 &  $-25.9$ & $-140.5$ &  $-56.7$   \\
2009 & 274 &   79.4 &   82.3 &  287.7 &  $-64.9$ &  $-28.9$ &  $-26.4$   \\
2009 & 281 &   75.5 &   75.3 &  256.7 &   14.0 &   34.2 &   29.4   \\
2009 & 282 &   47.2 &   60.8 &   47.6 &   11.5 &  168.6 &  $-38.7$   \\
2009 & 288 &   34.2 &   58.6 &  217.9 &  $-51.5$ &  $-41.6$ &    8.4   \\
2009 & 304 &   30.1 &   55.6 &  177.7 &   $-5.0$ &  $-83.8$ &   54.7   \\
2009 & 335 &   64.2 &   52.5 &  171.3 &  $-43.8$ &  $-73.1$ &   16.4   \\
2010 &  24 &   73.6 &   54.3 &   97.2 &   34.3 &  179.7 &   10.6   \\
2010 &  45 &   70.0 &   61.5 &  174.7 &  $-21.2$ &  $-78.9$ &   38.6   \\
2010 &  50 &   71.7 &   64.5 &  227.9 &  $-21.5$ &  $-18.6$ &   30.7   \\
2010 &  52 &   52.1 &   72.9 &  258.1 &  $-44.9$ &  $-17.0$ &   $-3.3$   \\
2010 &  72 &   43.3 &   66.9 &  278.8 &    7.9 &   38.2 &    7.2   \\
2010 & 121 &   43.6 &   82.0 &  122.7 &  $-70.7$ &  $-76.3$ &  $-19.3$   \\
2010 & 148 &   52.2 &   74.8 &   89.2 &  $-12.0$ & $-142.2$ &  $-17.5$   \\
2010 & 182 &   15.4 &   54.7 &  197.8 &  $-20.0$ &  $-50.7$ &   42.6   \\
2010 & 193 &   69.6 &   58.4 &  149.2 &    5.5 & $-127.5$ &   43.2   \\
2010 & 194 &   70.9 &   53.8 &  277.2 &    6.7 &   36.4 &    8.1   \\
2010 & 196 &   73.2 &   52.3 &  303.7 &  $-68.1$ &  $-32.6$ &  $-32.8$   \\
2010 & 204 &   38.7 &   53.2 &  180.5 &  $-11.5$ &  $-75.9$ &   49.6   \\
2010 & 205 &   47.4 &   53.5 &  315.8 &  $-82.1$ &  $-49.3$ &  $-31.2$   \\
2010 & 223 &   39.0 &   56.1 &  250.2 &  $-73.6$ &  $-42.6$ &  $-17.5$   \\
2010 & 224 &   62.3 &   65.2 &  284.7 &  $-28.2$ &    8.1 &  $-13.9$   \\
2010 & 226 &   53.8 &   75.6 &  324.5 &   17.9 &   71.2 &  $-25.0$   \\
2010 & 235 &   32.0 &   60.3 &  216.1 &  $-66.5$ &  $-48.0$ &   $-5.3$   \\
2010 & 238 &   12.4 &   69.6 &  226.4 &  $-25.7$ &  $-22.6$ &   28.1   \\
2010 & 239 &   66.7 &   58.4 &  312.9 &  $-14.2$ &   33.1 &  $-33.0$   \\
2010 & 256 &   73.8 &   76.1 &  131.9 &  $-15.5$ & $-118.9$ &   17.1   \\
2010 & 277 &   31.1 &   73.7 &   12.3 &  $-40.7$ &  $-55.3$ &  $-76.5$   \\
2010 & 284 &   48.6 &   89.1 &  218.8 &  $-70.8$ &  $-48.7$ &   $-9.7$   \\
2010 & 295 &   27.8 &   58.0 &    8.4 &  $-61.5$ &  $-53.3$ &  $-55.5$   \\
2010 & 310 &   45.4 &   53.1 &  118.1 &    8.5 & $-147.9$ &   17.4   \\
2010 & 311 &   58.4 &   70.5 &   64.2 &  $-46.5$ & $-107.2$ &  $-45.5$   \\
2010 & 319 &   11.4 &   55.0 &  118.6 &  $-37.4$ & $-107.2$ &   $-4.8$   \\
2010 & 320 &   29.0 &   54.3 &   80.2 &  $-64.1$ &  $-86.2$ &  $-34.1$   \\
2010 & 320 &    5.1 &   68.7 &  121.1 &  $-30.6$ & $-111.9$ &     0.4   \\
2010 & 342 &   40.5 &   54.6 &  170.9 &  $-43.7$ &  $-73.4$ &   16.4   \\
2010 & 347 &   24.6 &   54.9 &  231.9 &  $-56.6$ &  $-36.7$ &     0.0   \\
2010 & 348 &   33.8 &   54.4 &  179.7 &  $-68.6$ &  $-61.9$ &   $-6.2$   \\
2010 & 364 &   22.2 &   68.0 &  167.0 &  $-31.2$ &  $-81.8$ &   26.6   \\
2011 &  19 &   43.8 &   69.4 &  268.5 &  $-15.7$ &   12.4 &    5.1   \\
2011 &  26 &   25.0 &  100.1 &  150.1 &  $-10.3$ & $-110.9$ &   34.1   \\
2011 &  35 &   71.5 &   54.0 &  185.4 &  $-24.6$ &  $-65.6$ &   37.8   \\
2011 &  38 &   33.8 &   58.2 &   33.4 &  $-31.7$ & $-127.8$ &  $-71.5$   \\
2011 &  41 &   59.2 &   52.0 &  125.5 &  $-59.2$ &  $-86.0$ &  $-12.5$   \\
2011 &  45 &   25.5 &   62.7 &  215.5 &  $-10.1$ &  $-23.5$ &   46.8   \\
2011 &  49 &   39.3 &   60.3 &  239.4 &    3.9 &   13.8 &   39.9   \\
2011 &  75 &   60.5 &   71.1 &  230.3 &    1.5 &    3.8 &   45.9   \\
2011 &  86 &   59.4 &   56.2 &  160.3 &   $-3.$1 & $-108.3$ &   46.4   \\
2011 & 106 &   78.2 &   81.4 &  308.8 &   16.1 &   59.9 &  $-14.3$   \\
2011 & 111 &   65.6 &   69.7 &   30.3 &    3.8 &  154.2 &  $-54.8$   \\
2011 & 113 &   71.5 &   54.8 &  295.1 &  $-27.6$ &   12.2 &  $-22.3$   \\
2011 & 119 &   53.0 &   67.3 &  255.4 &   $-5.1$ &   14.8 &   21.6   \\
2011 & 120 &   49.8 &   72.1 &   84.9 &   14.4 & $-168.3$ &   $-8.7$   \\
2011 & 132 &   10.6 &   56.8 &   39.5 &  $-29.9$ & $-134.1$ &  $-66.5$   \\
2011 & 136 &   54.1 &   64.9 &  333.8 &  $-79.2$ &  $-48.7$ &  $-35.3$   \\
2011 & 162 &   72.4 &   55.9 &  132.8 &   12.9 & $-145.5$ &   32.4   \\
2011 & 203 &   29.9 &   77.9 &  120.8 &  $-56.3$ &  $-89.8$ &  $-13.2$   \\
2011 & 207 &   65.0 &   56.4 &  344.5 &  $-19.9$ &   42.3 &  $-63.1$   \\
2011 & 215 &   34.5 &   68.3 &  245.4 &  $-18.2$ &   $-2.8$ &   21.8   \\
2011 & 221 &    2.9 &   70.8 &  139.8 &  $-35.8$ &  $-98.2$ &    9.6   \\
2011 & 240 &   46.5 &   58.8 &  219.1 &  $-41.9$ &  $-36.9$ &   16.8   \\
2011 & 252 &   24.5 &   80.9 &  283.7 &  $-28.6$ &    7.4 &  $-13.2$   \\
2011 & 294 &   31.8 &   75.6 &   77.2 &  $-41.0$ & $-114.4$ &  $-36.1$   \\
2011 & 307 &   40.7 &   52.4 &  313.5 &  $-16.6$ &   30.7 &  $-34.4$   \\
2011 & 309 &   38.8 &   63.3 &   26.1 &  $-32.2$ & $-120.2$ &  $-77.4$   \\
2011 & 316 &   31.0 &   70.2 &    4.6 &  $-37.9$ &  $-26.2$ &  $-77.2$   \\
2011 & 318 &   36.7 &   57.2 &  148.8 &  $-13.0$ & $-109.6$ &   31.4   \\
2011 & 360 &   36.1 &   67.4 &  305.5 &  $-34.5$ &    7.6 &  $-32.7$   \\
2011 & 361 &   47.6 &   92.8 &  343.4 &  $-71.6$ &  $-44.9$ &  $-42.6 $  \\
2011 & 364 &   51.7 &   64.8 &  207.1 &  $-29.1$ &  $-42.4$ &   32.1   \\
2012 &  12 &   31.8 &   62.4 &   15.3 &   $-3.6$ &  129.0 &  $-66.3$   \\
2012 &  52 &   23.8 &   66.1 &   33.2 &  $-59.0$ &  $-75.3$ &  $-55.2$   \\
2012 &  81 &   47.3 &   99.0 &  309.4 &  $-66.8$ &  $-31.5$ &  $-35.2$   \\
2012 & 103 &   67.5 &   70.4 &  154.0 &  $-46.3$ &  $-83.1$ &    8.6   \\
2012 & 109 &   25.9 &   62.6 &   37.8 &  $-39.5$ & $-110.0$ &  $-65.9$   \\
2012 & 132 &   62.3 &   58.5 &  189.0 &   $-5.1$ &  $-64.1$ &   57.6   \\
2012 & 154 &   65.8 &   58.7 &   37.0 &  $-75.8$ &  $-64.6$ &  $-39.9$   \\
2012 & 155 &   64.3 &   60.0 &  245.4 &  $-30.9$ &  $-12.7$ &   13.3   \\
2012 & 162 &   58.5 &   83.8 &   26.8 &  $-24.8$ & $-154.6$ &  $-77.3$   \\
2012 & 183 &   59.8 &   61.8 &  259.8 &  $-32.7$ &   $-6.2$ &    2.7   \\
2012 & 189 &   31.4 &   61.1 &   18.7 &  $-42.5$ &  $-72.9$ &  $-73.9$   \\
2012 & 193 &   65.5 &   54.4 &  342.9 &   $-6.5$ &   63.4 &  $-54.$8   \\
2012 & 206 &   61.6 &   56.8 &  310.6 &  $-83.1$ &  $-50.0$ &  $-30.2$   \\
2012 & 211 &   50.0 &   58.7 &  177.2 &   12.5 & $-105.1$ &   69.3   \\
2012 & 301 &   38.5 &   53.3 &   56.3 &   $-3.2$ & $-169.2$ &  $-42.1$   \\
2012 & 332 &   48.1 &   71.1 &  227.6 &   11.9 &   14.7 &   54.0   \\
2013 &  11 &   17.0 &   55.7 &  217.1 &  $-24.5$ &  $-30.5$ &   33.3   \\
2013 &  27 &   26.5 &   62.7 &  200.9 &  $-34.6$ &  $-49.6$ &   27.8   \\
2013 &  27 &   47.6 &   70.7 &   56.6 &  $-67.8$ &  $-77.6$ &  $-41.7$   \\
2013 &  31 &   67.3 &   53.2 &  314.9 &  $-67.3$ &  $-32.8$ &  $-37.1$   \\
2013 &  36 &   74.7 &   73.6 &  267.5 &  $-68.3$ &  $-34.$8 &  $-19.7$   \\
2013 &  52 &   60.7 &   71.9 &   73.7 &  $-20.5$ & $-139.8$ &  $-34.4$   \\
2013 &  70 &   41.9 &   53.9 &  154.3 &  $-15.8$ & $-102.7$ &   33.1   \\
2013 & 119 &   61.5 &   62.1 &  138.6 &   26.1 & $-158.8$ &   41.9   \\
2013 & 132 &   59.3 &   57.3 &  357.0 &  $-81.1$ &  $-54.1$ &  $-35.7$   \\
2013 & 134 &   44.9 &   85.3 &  123.4 &   $-6.2$ & $-131.7$ &   15.1   \\
2013 & 144 &   49.8 &   54.3 &   33.3 &  $-39.0$ & $-107.2$ &  $-69.2$   \\
2013 & 163 &   44.6 &   52.2 &     0.4 &  $-68.1$ &  $-50.1$ &  $-48.3$   \\
2013 & 175 &   50.6 &   58.9 &  211.1 &   15.0 &    1.0 &   69.1   \\
2013 & 190 &   57.3 &   68.8 &   64.7 &  $-70.1$ &  $-77.0$ &  $-38.0$   \\
2013 & 191 &    8.8 &   67.3 &  308.1 &  $-39.5$ &    2.1 &  $-35.7$   \\
2013 & 222 &   63.4 &   61.5 &  240.3 &  $-68.9$ &  $-41.3$ &  $-12.1$   \\
2013 & 224 &   47.9 &   63.4 &  345.4 &   $-9.0$ &   62.7 &  $-58.3$   \\
2013 & 247 &   54.7 &   84.8 &  154.6 &  $-46.9$ &  $-82.4$ &    8.3   \\
2013 & 249 &   30.0 &   55.5 &  160.4 &  $-34.8$ &  $-85.2$ &   20.9   \\
2013 & 249 &   55.0 &   65.4 &   92.1 &  $-64.1$ &  $-86.4$ &  $-28.9$   \\
2013 & 281 &   65.1 &   58.5 &  327.5 &  $-25.1$ &   25.3 &  $-49.4$   \\
2013 & 297 &   39.0 &   73.0 &  163.8 &  $-74.1$ &  $-64.9$ &  $-13.1$   \\
2013 & 302 &   49.4 &   54.6 &  298.7 &    8.8 &   48.2 &   $-9.8$   \\
2013 & 319 &   62.0 &   54.4 &  284.5 &  $-37.6$ &   $-1.0$ &  $-17.3$   \\
2013 & 320 &   22.2 &   52.9 &  286.8 &  $-55.0$ &  $-18.3$ &  $-24.1$   \\
2013 & 329 &   29.2 &   63.6 &  182.3 &  $-14.3$ &  $-72.3$ &   47.3   \\
2013 & 332 &   31.1 &   65.2 &  241.6 &  $-53.5$ &  $-30.5$ &   $-1.0$   \\
2013 & 352 &   51.4 &   72.5 &   91.4 &  $-60.6$ &  $-90.4$ &  $-28.9$   \\
2013 & 364 &   60.2 &   53.2 &  198.8 &  $-63.9$ &  $-54.5$ &   $-1.2$   \\
2014 &   8 &   57.9 &   60.0 &   72.8 &  $-73.5$ &  $-74.4$ &  $-34.3$   \\
2014 &  30 &   60.8 &   74.5 &  189.9 &  $-32.7$ &  $-60.0$ &   30.1   \\
2014 &  32 &   12.8 &   54.6 &  186.7 &  $-24.9$ &  $-64.1$ &   37.6   \\
2014 &  49 &   41.7 &   54.9 &    2.3 &  $-49.2$ &  $-39.7$ &  $-66.4$   \\
2014 &  59 &   25.9 &   60.2 &  239.5 &  $-49.2$ &  $-28.7$ &    3.0   \\
2014 &  64 &   66.7 &   63.6 &   45.2 &  $-65.8$ &  $-75.6$ &  $-46.4$   \\
2014 &  65 &   58.5 &  118.3 &  340.6 &   12.0 &   80.1 &  $-39.9$   \\
\enddata

\end{deluxetable}

\end{document}